\documentclass[PRL,preprintnumbers,floatfix,aps,notitlepage,nofootinbib,twocolumn,showpacs,amssymb]{revtex4-1}
\usepackage{amsmath,amsfonts,bm}
\usepackage{graphicx}
\usepackage{cancel}
\usepackage[colorlinks, linkcolor={red},citecolor={blue}]{hyperref}
\usepackage{xcolor}
\usepackage{color}

\begin{document}
\preprint{FQSP-25-1, YITP-25-70 
}
\title{An analytic model for a total process of gravitational collapse: \\From star to Schwarzschild black hole}
\author{Sinya Aoki}
\email[]{saoki@yukawa.kyoto-u.ac.jp}
\affiliation{Fundamental Quantum Science Program (FQSP), TRIP Headquarters, RIKEN,
	Hirosawa 2-1, Wako, Saitama 351-0198, Japan}
\affiliation{Center for Gravitational Physics and Quantum Information,
	Yukawa Institute for Theoretical Physics, Kyoto University,
	Kitashirakawa Oiwakecho, Sakyo-Ku, Kyoto 606-8502, Japan}
	
\author{Jorge Ovalle}
\email[]{jorge.ovalle@physics.slu.cz}
\affiliation{Research Centre for Theoretical Physics and Astrophysics,
	Institute of Physics, Silesian University in Opava, CZ-746 01 Opava, Czech Republic 
}
\affiliation{Universidad de Tarapac\'a, Avenida Luis Emilio Recabarren 2477, Iquique, Chile
}

\begin{abstract}
	We present an exact analytical model describing a complete gravitational collapse of matter from horizonless initial conditions to black hole formation, tracing the full evolution of the horizon $H(t)$ from its formation at microscopic scales to macroscopic stabilization. The solution reveals two main stages: (i) a dynamical horizon growth, where an apparent horizon $H(t)$ emerges at a critical time $t_b$ until reaching its final size $h=2{\cal M}$, demonstrating how trapped surfaces form dynamically in finite time, and (ii) a {naked-}singularity resolution, where an integrable Ricci curvature singularity ($R^\mu{}_\nu \sim r^{-2}$) develops at $r=0$, but remains causally hidden by the horizon growth, preserving weak cosmic censorship without exotic matter. The model could offer a framework to study the quantum-to-classical transition ($H(t) \sim \ell_{\rm Planck}$).
\end{abstract} 
\maketitle
%
%
%
\section{Introduction}

The gravitational collapse of matter and subsequent black hole (BH) formation has remained a central focus of theoretical and observational physics since Oppenheimer and Snyder's seminal work~\cite{Oppenheimer:1939ue}. †In particular, in spherically symmetric systems, efforts to provide an exact analytical description of BH formation encounter considerable obstacles~\cite{Bondi:1947fta,Christodoulou:1984mz,Joshi:1993zg,Shibata:1999va,Joshi:2001xi,Giambo:2003fd,Goswami:2003bs,Lasky:2006mg,Mosani:2020ena,BenAchour:2020gon}. (see Ref.~\cite{Joshi:2011rlc} for a comprehensive review, and Refs.~\cite{Greenwood:2008ht,Wang:2009ay,Kiefer:2019csi,Blanchette:2020kkk,Piechocki:2020bfo,Kelly:2020lec,Husain:2021ojz,Husain:2022gwp,Casadio:2022pla,Cipriani:2024nhx,Casadio:2026tmd} for quantum approaches). This field has witnessed landmark developments, from Penrose's foundational singularity theorems~\cite{Penrose:1964wq}, to the groundbreaking imaging of BH shadows by the Event Horizon Telescope collaboration~\cite{EventHorizonTelescope:2019dse,EventHorizonTelescope:2022wkp}. Despite these advances, fundamental questions persist regarding the dynamics of collapse and the final state configuration. Key unresolved issues include (i) Horizon formation: The precise mechanism of horizon emergence during collapse and (ii) Singularity resolution: The connection between non-singular cores and stability of interior horizons (e.g., Cauchy horizons)~\cite{Poisson:1989zz,Poisson:1990eh}. These questions bear directly on the validity of cosmic censorship conjectures~\cite{Penrose:1969pc,Penrose1979,Wald:1984rg} and the predictability of physics inside BHs.

Building upon these considerations, particularly the potential description of BH interiors with structures more complex than singularities, a recent extension of the Schwarzschild solution has been proposed~\cite{Ovalle:2024wtv,Ovalle:2025pue}, exhibiting three remarkable characteristics: (i) Requires no additional parameters beyond the BH mass ${\cal M}$ (i.e., retains zero primary hair), (ii) Excludes exotic matter by construction, and (iii) Remains fully within the framework of classical general relativity (see Ref.~\cite{Ovalle:2026lxb} for a detailed analysis of the near-core region and the so called ``Minkowski breaking'').

This demonstrates that the solution faithfully describes the Schwarzschild geometry prior to complete gravitational collapse, without introducing additional degrees of freedom.  Crucially, if we impose the weak cosmic censorship conjecture~\cite{Penrose:1969pc} to preclude naked singularities, the event horizon must necessarily form prior to the emergence of a central singularity. This implies that a portion of the total mass ${\cal M}$, already enclosed by the horizon, remains in a pre-singularity state during collapse. The Schwarzschild revisited solutions explicitly realize this scenario, revealing an unexpected richness in possible interior structures for Schwarzschild BHs. These solutions hold significance beyond their simplicity and novelty, as they provide a powerful framework for probing gravitational collapse evolution~\cite{Aoki:2024dyr,Ovalle:2025pue} (see Refs.~\cite{Casadio:2024fol,Casadio:2025pun} for Cosmological models and Ref.~\cite{Casadio:2026tmd} for a quantum model).

In Ref.~\cite{Aoki:2024dyr}, we analyzed gravitational collapse mo\-dels departing from the revisited Schwarzschild BH. The present work extends this framework to describe the complete collapse process: from an initial horizon-free matter configuration of total mass ${\cal M}_0$ to an intermediate BH state with horizon $h=2{\cal M}$, where we assume perfect matter collapse (${\cal M} = {\cal M}_0$). While realistic collapse typically leaves residual matter (e.g., accretion disks) with ${\cal M} < {\cal M}_0$, we adopt a simplified scenario where all matter collapses into the BH. This simplified model isolates the fundamental physics of horizon formation while maintaining analytical tractability. Most importantly, it establishes an exact framework describing the continuous gravitational evolution from an initial seed configuration, a micro BH, to its final macroscopic state.

The paper is organized as follows: in section ~\ref{sec2}, we first briefly review the previous study on spherically symmetric static BHs containing only integrable singularities~\cite{Lukash:2013ts,Ovalle:2023vvu,Arrechea:2025fkk} and their eventual gravitational collapse to the Schwarzschild BH proposed in Ref.~\cite{Aoki:2024dyr} .
In section~\ref{sec3}, 
we first resolve a pathology present in Model II of Ref.~\cite{Aoki:2024dyr}, where the BH surface becomes acausal at very late times, shortly before the formation of the Schwarzschild BH.
We then extend the analytical model of gravitational collapse to a process encompassing the collapse from the birth of an infinitesimal horizon BH (micro BH) to one of maximum size. 
We add a further extension showing a mechanism that prevents naked singularities prior to the emergence of the micro BH.
Finally we summarize our conclusions in section ~\ref{con}. In appendix~\ref{app:EnergyCondition}, we present a detailed analysis of the energy conditions.


\section{BH with  integrable singularities and its collapse to the Schwarzschild BH}
\label{sec2}
\subsection{BH with  integrable singularities }
In order to be as self-contained as possible, we will now briefly describe the interior of a spherically symmetric static BH (for all details see Ref.~\cite{Ovalle:2024wtv}). 
For the later convenience, we write the original metric~\cite{Ovalle:2024wtv} in the Eddington-Finkelstein form as 
 \begin{equation}
	ds^{2}
	= -(1+u) dt^2 - 2u dt dr +(1-u) dr^2 + r^2 d\Omega^2
	\ ,
	\label{metric}
\end{equation}
where $d\Omega^2= d\theta^2 +\sin^2\theta d\phi^2$ and 
\begin{equation}
u(r) = -{2 m(r)\over r}.
\end{equation}
The $r$-dependent mass function is given by
\begin{eqnarray}
m(r) &=&\left\{
\begin{array}{ll}
h f( x),  & x:= \dfrac{r}{h} \le 1,     \\
\\
 {\cal M}:= \dfrac{h}{2}, & x > 1 ,    \\
\end{array}
\right.
\label{mass}
\end{eqnarray}
where ${\cal M}$ and $h$ are the BH mass and its horizon, and the function $f(x)$  is chosen to satisfy the following conditions, 
\begin{equation}
f(1)=\dfrac{1}{2}, \ f^\prime(1)=f^{\prime\prime}(1)=0,\ f(0)=0,\ f^\prime(0)=1,
\label{function_f}
\end{equation}
whose explicit solutions are given in Ref.~\cite{Ovalle:2024wtv}.
The simplest one, for example, denotes
\begin{equation}
f(x)=f_0(x) :=x - x^3 +{x^4\over 2}.
\label{simple}
\end{equation}

The constant $t$ hypersurface is spacelike even inside the horizon in this coordinate,
since its normal vector is given by $n_\mu =\delta^t_\mu$ [equivalently $n^\mu= -(1-u)\delta^\mu_t -u\delta^\mu_r$], and thus
timelike as $ n^\mu n_\mu = -(1-u) < 0 $ for all $r$.
Therefore we can regard $t$ as an appropriate time coordinate.

The metric in Eq.~\eqref{metric} coincides with the Schwarzschild solution~\cite{schwarzschild1916b} outside the horizon $r > h$. However, the interior solution ($r \le h$)  differs significantly due to the presence of a non-vacuum matter distribution, resulting in a curvature singularity at the origin $r=0$. This singularity is weaker than that of a Schwarzschild BH: the metric remains finite with $u(0)= -2 h$, and the Ricci tensor exhibits an integrable divergence $R^\mu{}_\nu \propto r^{-2}$ near the origin, as will be shown later.

At the horizon, the mass function $m(r)$ exhibits not only continuity but also smoothness up to second order, with
\begin{equation}
	m(h) = \mathcal{M}, \quad m'(h) = m''(h) = 0,
	\label{dmass}
\end{equation}
as guaranteed by the conditions imposed on Eq.~\eqref{function_f}.

As previously noted, within the horizon $r\le h$, the energy-momentum tensor (EMT) is non-vanishing and takes the form
\begin{eqnarray}
	T^t{}_t &=& T^r{}_r = -\frac{2 m^\prime}{\kappa r^2}, \quad T^\theta{}_\theta = T^\phi{}_\phi = -\frac{m^{\prime\prime}}{\kappa r},
\end{eqnarray}
where $\kappa = 8\pi G_N$  denotes the gravitational coupling constant. Due to the smoothness properties specified in Eq.~\eqref{dmass}, all components of the EMT vanish identically at the horizon $r=h$, ensuring continuity across this boundary. The Ricci tensor $R^\mu{}_\nu=\kappa T^\mu{}_\nu$ inherits its behavior from the EMT, that is, with $m^\prime\sim1$ and $m^{\prime\prime}\sim r^{-1}$ near the origin, it exhibits an $r^{-2}$ divergence. Importantly, this matter distribution satisfies the strong and weak {(and therefore null)}  energy conditions~\cite{Ovalle:2024wtv}. Consequently, interior matter is physically admissible (there is no exotic matter) and no additional horizons form within $ r\le h$, preserving the causal structure.

\subsection{Gravitational collapse to the Schwarzschild BH}
To describe  gravitational collapse, we introduce the time dependence in the mass function as $m(r) \to m(r,t)$, which satisfies
$m(r,0)=m(r)$ and $m(r,\infty) = {\cal M}$:
The BH with integrable singularity characterized by $m(r)$ in Eq.~\eqref{mass} at $t=0$ collapses to the Schwarzschild BH with mass ${\cal M}$ at $t=\infty$.
In this modification, the EMT reads~\cite{Aoki:2020nzm,Aoki:2022gez} 
\begin{eqnarray}
T^{t}{}_{t}&=&-\frac{2m_r}{\kappa r^2}:=-A\, , \label{Ttt}\\
T^{r}{}_{r} &=&-\frac{2}{\kappa r^2}\left(m_r-2 m_t\right):= -A+2B\, , \label{Trr}\\
T^\theta{}_{\theta}&=&T^\phi{}_{\phi}=-\frac{1}{\kappa r}\left(m_{rr}-2 m_{rt}+m_{tt}\right):=-C\, ,\label{Tpp}\\
T^t{}_{r}&=&-T^r{}_{t}=-\frac{2m_t}{\kappa r^2}:= -B\, , \label{Ttr}
\label{EMT}
\end{eqnarray}
where 
\begin{equation}
m_a(r,t) := \frac{\partial\,m (r,t)}{\partial\,a}, \quad m_{ab}(r,t) :=\frac {\partial^2\,m (r,t)}{\partial\,a \partial\, b}
\end{equation}
for $a,b= r,t$.
We here introduce $A,B,C$ for later uses.

In the previous work~\cite{Aoki:2024dyr}  we studied the dynamical BH model with mass function for $0\le t < \infty$:
\begin{eqnarray}
	m(r,t) &=& \left\{
	\begin{array}{ll}
		2{\cal M} f(x(r,t)), & \displaystyle  x(r,t) \le 1,     \\
		\\
		{\cal M}, & x(r,t)  > 1 ,    \\
	\end{array}
	\right.
	\label{mass2}
\end{eqnarray}
where the dimensionless radial coordinate $x(r)$, introduced in~\eqref{mass}, has been promoted to a time {dependent one} as
\begin{equation}
	x(r,t) := \dfrac{r}{R(t)}, \quad R(t) := h e^{-a t}, \ a:={\alpha\over h}
	\label{eq:t-dep1}
\end{equation}

In Fig. 4 of Ref.~\cite{Aoki:2024dyr},  profiles of $m(r,t)$ were shown as a function of $r$. As $t$ increases, while the  horizon keeps its size $h$, a vacuum region inside the horizon develops {between}  $h$ {and} $h e^{-\alpha t/h}$,
and the mass function $m(r,t)$ approaches the profile of the constant $m(r)={\cal M}$.

Since $m(0,{}^\forall t)=0$  by construction, the matter distribution never reaches the central singularity $r=0$ at any finite time $t$. As $t$ increases, the mass function $m(r,t)$ uniformly converges to the Schwarzschild solution everywhere except at the origin, where the vanishing mass $m(0,t)=0$ persists. Therefore we obtain
\begin{eqnarray}
\lim_{t\to\infty} m(r,t) ={\cal M} \theta (r),
\label{eq:SBH_mass}
\end{eqnarray}
which is nothing but a mass function of the Schwarzschild BH with a regularized singularity at $r=0$ by the step function 
$\theta(r)$ satisfying $\theta(r)=1$ for $ r>0$ with $\theta(0)=0$~\cite{Aoki:2020prb}. 

Accordingly the scalar curvature and Kretschmann scalar behave respectively as
\begin{eqnarray}
	\label{scalars2}
	R_{\rm Ricci}&=&\left\{
\begin{array}{ll}
  2(A-B-C), & x(r, t) \le 1 ,    \\
  \\
  0, &  x(r,t) > 1 ,   \\
\end{array}
\right.
\label{ricci}
\\
	\label{krets2}
R_{\mu\nu\sigma\rho}R^{\mu\nu\sigma\rho} &=&
\left\{
\begin{array}{ll}
  \dfrac{16e^{2\alpha t/h}}{r^4}+\cdots, & x(r,t) \le 1,     \\
  \\
  \dfrac{12 h^2}{r^6}, &  x(r,t) > 1,    \\
\end{array}
\right.
\end{eqnarray}
which shows a curvature singularity at $r=0$ that is weaker than the Schwarzschild BH's singularity for finite $t$, thank to the mass function property $m(0,t)=0$. Since the region of $r$ satisfying $x(r,t)\le 1$ shrink to $r=0$ as $t\to\infty$, 
$R_{\rm Ricci}=0$  ($r \not= 0$) and $R_{\mu\nu\sigma\rho}R^{\mu\nu\sigma\rho} =12 h^2 r^{-6}$ ($r \not= 0$) in this limit.

Since
\begin{eqnarray}
m_a(r,t) &=& h f^\prime \left( x(r,t) \right) x_a(r,t), 
\label{d_mass} \\
m_{ab} (r,t) &=& h  f^\prime \left( x(r,t) \right) x_{ab}(r,t)\nonumber \\
 &+& h f^{\prime\prime} \left( x(r,t) \right) x_a(r,t) x_b(r,t) 
 \label{dd_mass}
\end{eqnarray}
at $x(r,t) \le 1$ for $a,b = r,t$, we see that $m_a(r,t) = m_{ab} (r,t) =0$ at $x(r,t) = 1$ for all $a,b$, thank to properties of $f(x)$ in Eq.~\eqref{function_f}.
Therefore all components of the EMT in Eq.~\eqref{EMT} vanish at $x(r,t)=1$, so that the EMT is continuous there.

This EMT satisfies the null energy condition (NEC) and the strong energy condition (SEC) at $\alpha \le \alpha_0=3.13422\cdots$, while  the weak energy condition (WEC) at $\alpha \le 1$~\cite{Aoki:2024dyr}. Therefore this collapse is described by physically reasonable matter.

\section{Model with a growing  horizon}
\label{sec3}
In the gravitational collapse model of Ref.~\cite{Aoki:2024dyr}, the BH horizon maintains a constant size $h$ while the interior matter distribution evolves dynamically. We now extend this analysis to the earlier collapse phase, where the {(apparent)} horizon forms from initially horizon-free configurations and grows to its final size $h$. 

{The complete gravitational collapse process occurs through three distinct phases:}

\begin{enumerate}
	\item {Pre-Horizon Phase} ($t_a \leq t < t_b$):  
	The collapse initiates at $t_a$ with a spherically symmetric, horizon-free matter distribution of radius $R(t_a) > 2h$, where $h$ denotes the final horizon size. The exterior region [$r > R(t)$] remains Schwarzschild vacuum. As matter collapses inward, $R(t)$ monotonically decreases while the vacuum region expands, culminating at $t_b$ when $R(t_b) = 2h$.
	
	\item {Horizon Growth Phase} ($t_b \leq t \leq t_c$):  
	A dynamical (micro) horizon $H(t)$ emerges at $t_b$ ($H(t_b) = 0$) and grows, partitioning spacetime into:
	\begin{itemize}
		\item A BH interior ($0 < r \leq H(t)$)
		\item An exterior matter shell ($H(t) < r \leq R(t)$)
		{\item A Schwarzschild vacuum at $r > R(t)$} 
	\end{itemize}
	The phase concludes at $t_c$ when $H(t_c) = R(t_c) = h$, with all matter fully contained within the horizon.
	
	\item Final Phase ($t_c< {t \le t_d}$):  
	As  in the case of  Ref.~\cite{Aoki:2024dyr}, the system evolves toward a Schwarzschild BH with mass $\mathcal{M} = h/2$ as $t \to {t_d}$, {where $t_d=\infty$ in Ref.~\cite{Aoki:2024dyr}.}
\end{enumerate}
We emphasize that the horizon size $H(t)$ evolves continuously from the formation ($t_b$) to its maximum value ($t_c$), i.e., from a micro BH to a macroscopic configuration.

\subsection{Causality problem and improved model}

A problem we find for the model II in  Ref.~\cite{Aoki:2024dyr} is  that the time dependence in Eq.~\eqref{eq:t-dep1} violates causality.
To see this, let us consider  the boundary hypersurface between the matter existing region and vacuum ({\it i.e.} the surface of the collapsing star), given by $r= R(t)$ ( $x(r,t)=1$).  The vector normal to this hypersurface becomes $N_\mu (t) dx^\mu = a R(t) d t + dr$, whose norm is given by
 \begin{equation}
N^2(t):= g^{\mu\nu} N_\mu N_\nu =( 1-a R ) \left[ 1+ u +(1-u) a R \right]\nonumber
\end{equation}
with $u(t) = - h/R(t)$.
Since $N^2(t)\to -\infty$ as $t\to\infty$, the hypersurface becomes space-like ($N^2(t) < 0$ ) at $ t > t_0$ for some finite $t_0$.
Therefore, matters on the hypersurface  move faster than light violating causality, despite the fact that the matter EMT satisfies the energy conditions.\\

We therefore modify the time dependence in order to satisfy causality.
For a general non-negative function $R(t)$, $N^2(t)$ becomes
\begin{eqnarray}
N^2(t)=\frac{1+\dot R}{R}[R-h-(R+h)\dot R],\ \dot R := {d R\over d t},
\end{eqnarray}
and  conditions required for $R(t)$ to satisfy are summarized as follows.
\begin{enumerate}
\item The total process of collapse starts at $t=t_a$, when $\dot R(t_a)=0$.  
\item  The collapse ends at $t=t_d$, when  $R(t_d)=0$. Only the Schwarzschild BH remains at $t \ge t_d$.
\item   $R(t)$ is monotonically decreasing: $\dot R \le 0$ for all $ t_a\le  t\le t_d$.
\item   The apparent horizon starts forming at $t=t_b$, so that  $R(t_b)=2h$. Without  loss of generality, we take $t_b=0$. {\it i.e.}, $R(0)=2h$, using a freedom to shift the origin of $t$.
\end{enumerate}
Under these conditions, 
we determine  $R(t)$, in order to satisfy 
that $N^2(t) \ge 0$ for $t_a\le t \le t_d$, so that the hypersurface defined by $r=R(t)$ is time-like or null.\\

Using the condition 2, $R(t_d)=0$, we write $R(t)=(t_d-t) g(t)$ where $g(t)$ is expanded in terms of $T:= t_d-t$. 
Thus using the condition 3 at $t=t_d$, we have $\dot R(t_d) = - g(t_d) \le 0$, leading to  $g(t_d) \ge 0$.

With the condition 4,  we have 
\begin{equation}
N(0)= \left(1+\dot R(0)\right)\frac{1-3\dot R(0)}{2}  \ge 0 \to  -1 \le \dot R(0) \le 0.
\label{eq:N0}
\end{equation}

Since the causality requires 
\begin{equation}
\lim_{t\to t_d} N^2(t) =-{h (1-g(t_d))^2\over g(t_d) T} \ge 0,
\end{equation}
we need to take $g(t_d)=1$. Therefore we can write
\begin{equation}
R(t) = T \Bigl( 1-\sum_{n=1}^\infty b_nT^n\Bigr),
\label{eq:generalR}
\end{equation}
and we adjust $b_n$ to satisfy $N^2(t) \ge 0$ for $t_a\le t\le t_d$.

\bigskip

The simplest model is given at $T^2$ as
\begin{equation}
R(t) =T( 1- b_1 T ) \to \dot R(t) = -1 + 2b_1 T.
\label{eq:simpleR}
\end{equation}
The condition 4, $R(0)=2h$, leads to $b_1=\dfrac{t_d-2h}{t_d^2}$,
while Eq.~\eqref{eq:N0} implies $0\le b_1\le \dfrac{1}{2t_d}$. Combining these we obtain the constraint for $t_d$ as
\begin{equation}
2h \le  t_d \le 4h.
\end{equation}

The condition 1, $\dot R(t_a)=0$, implies 
$ t_a = \dfrac{t_d(t_d-4h)}{2(t_d-2h)} < 0$, equivalently $T_a:=T(t_a) = \dfrac{1}{2 b_1}$.\\

We now write 
\begin{equation}
N^2(t) ={4b_1^3 T\over (1-b_1T )}  f(T), \ f(T):= (T -T_+) (T -T_-),
\label{eq:N2}
\end{equation}
where
\begin{equation}
0< T_-:={1-\sqrt{b_1 h}\over b_1} <  T_+:={1+\sqrt{b_1 h}\over b_1}.
\end{equation}
Therefore $f(T) \ge 0$ for $0  \le T \le T_a$ ($T(t_d)=0$), since
\begin{equation}
T_a - T_-  =  \frac{2\sqrt{b_1 h} -1}{2 b_1} < 0.
\end{equation}
Furthermore $1- b_1 T \ge 1-b_1 T_a =1/2$ for  $ 0 \le T \le T_a$.
Therefore eq.~\eqref{eq:N2} concludes  that $N^2(t) \ge 0$ is satisfied for $t_a \le t \le t_d$.\\

The total time for the gravitational collapse is given by $ t_d-t_a =  \dfrac{t_d^2}{2(t_d-2h)} \ge 4 h$ for $ 2h < t_d \le 4h$,
which is proportional to $h$.
Larger the total mass, longer the total time to complete the collapse. \\

As shown in the appendix, the EMT  satisfies NEC, WEC and SEC  at least for the simplest model.

Hereafter, we employ the simplest model, Eq.~\eqref{eq:simpleR}, in our analysis.

\subsection{BH with a growing horizon}
 \begin{figure}
	\centering
	\includegraphics[width=0.50\textwidth]{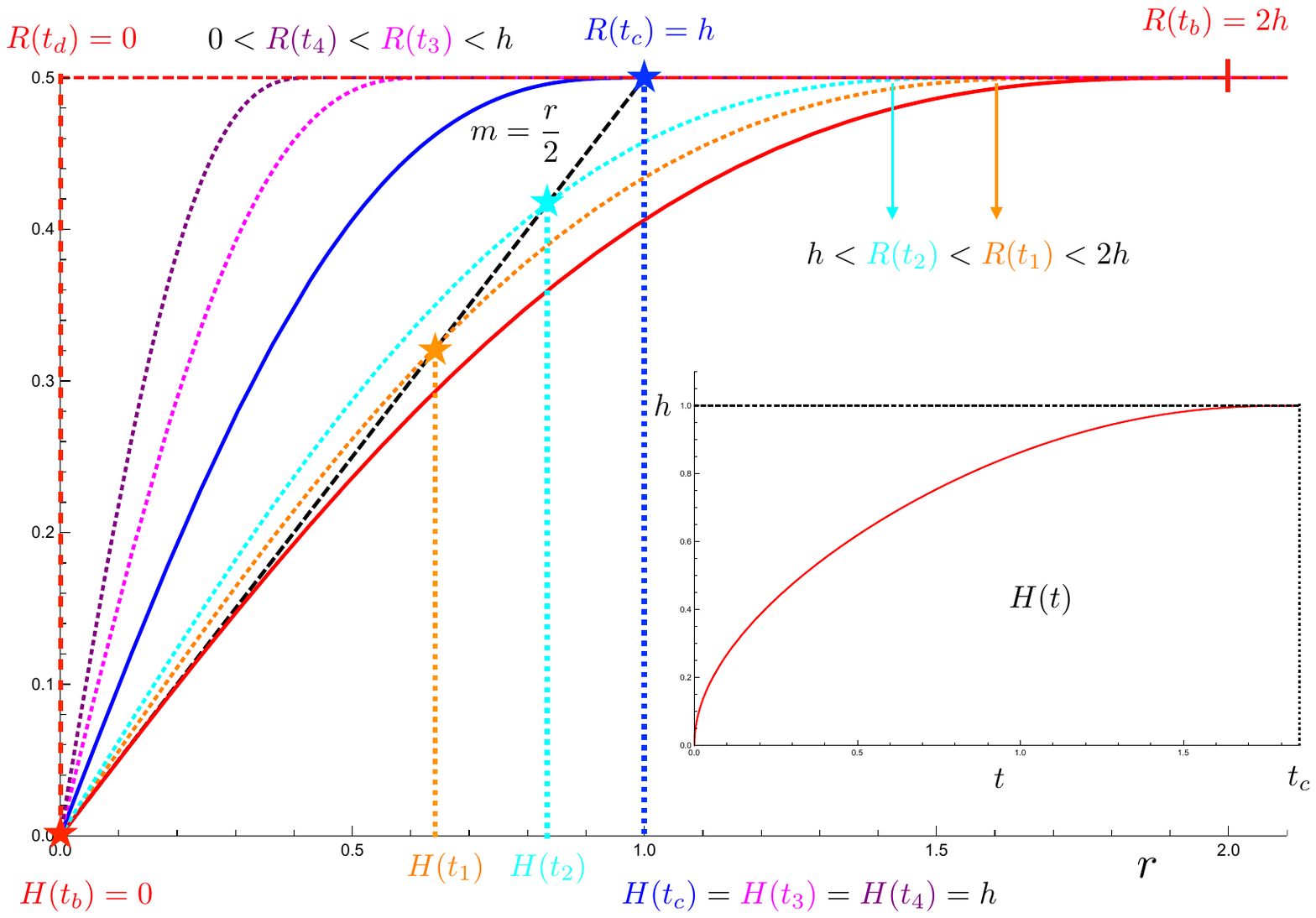}
	\caption{A profile of  $m(r,t)$ in Eq.~\eqref{mass2} with Eq.~\eqref{eq:simpleR} for the BH with the growing horizon as a function of $r$ at 
	 $t_b= 0$ (red solid line),  $t_1= 0.3 t_c$ (orange dotted line) ,  $t_2= 0.5 t_c$ (cyan dotted line),
 $t_c$  (blue solid line),  $t_3=0.4(t_d-t_c)+t_c $ (magenta dotted line),  $t_4=0.6(t_d-t_c)+t_c $ (purple dotted line) and $t_d$ (red dashed line), where $t_c$ is given in Eq.~\eqref{eq:tc}.
 The black dashed line represents $m= \dfrac{r}{2}$, whose intersection with the profile gives the horizon $H(t)$.
 We take  $f(x)=f_0(x)$, $t_d=3h$ and  $h=1$.
 Inset in the bottom-right shows the size of horizon $H(t)$ as a function $t$, which reaches to the maximum size $h$ at $t_c\sim\, 1.8541 $.  
 }
	\label{Fig3}
\end{figure}

The apparent horizon formation condition is determined by
\begin{eqnarray}
1+ u(r,t) = 0 \rightarrow m(r,t) = {1\over 2}r,
\end{eqnarray}
which admits solutions for $r$ if and only if the slope of $m(r,t)$ at $r=0$ satisfies $m_r(0,t) > \frac{1}{2}$, as visually demonstrated in Fig.~\ref{Fig3}. Note that this constraint follows necessarily from the boundary condition $m(0,t) = 0$, which holds throughout the collapse process.

Substituting the mass function from Eq.~\eqref{mass2} with Eq.~\eqref{eq:simpleR}, this condition becomes 
\begin{eqnarray}
m_r(0,t)=h f^\prime(0) x_r(0,t) = {h\over R(t)}  \ge {1\over 2}.
\end{eqnarray}
Consequently, the horizon first emerges at $t=t_b=0$ since $R(0)=2h$ by our choice of the origin for $t$.

During the formation phase $t_b \leq t <  t_c$, the horizon radius grows monotonically, ultimately reaching its final size $h$ at $t=t_c$.
The condition $R(t_c)=h$ is solved as
\begin{equation}
t_c= \frac{t_d}{2(t_d-2h)}\left[ \sqrt{t_d^2 -4h t_d +8h^2} +t_d-4h\right].
\label{eq:tc}
\end{equation}

In order to determinate the causal character of the apparent horizon $H(t)$, we proceed in the standard way. 
A vector normal to $H(t)$ is given by
\begin{eqnarray}
n_\mu dx^\mu = \frac{\dot R}{R}(1-x)^2(2x+1) dt +\frac{x(4-3x)}{2 R} dr, 
\end{eqnarray}
which leads to
\begin{eqnarray}
n^2&:=&n^\mu n_\mu = -2  \frac{\dot R}{R}(1-x)^2(2x+1)\nonumber \\
&\times& \left[  \frac{\dot R}{R}(1-x)^2(2x+1) -\frac{x(4-3x)}{2 R}\right].
\end{eqnarray}
Since $\dot R < 0$, it follows that $n^2 < 0$ for $0 \le x < 1$, while $n^2=0$ at $x=1$.
Therefore, the apparent horizon $H(t)$ is space-like for $ t_b= 0 \le t < t_c$, becoming nul at $t=t_c$.

Fig.~\ref{Fig3} shows a profile of  $m(r,t)$ for the BH with the growing horizon as a function of $r$ at several fixed {$t \ge t_b=0$}.
{Cases at $t < 0$ will be considered later.}
For $m(r,t)$, we take the simplest function $f(x)=f_0(x)$ in Eq.~\eqref{function_f} {with $t_d=3h$ and $h=1$. }
 The inset in the bottom-right of Fig.~\ref{Fig3} shows the time dependence of the horizon size $H(t)$ at $t\ge 0$,
 which reaches to the maximum size $h$ at $t=t_c$.

\begin{description}
		\item[Zero Horizon ($t =t_b= 0$)]  
	At $t=t_b=0$, the matter boundary $R(t_0)=2h$ causes the mass profile (red solid line) to intersect the horizon condition $m=\frac{r}{2}$ (black dashed line)  at $r=0$ {(red star)}, marking the birth of an infinitesimal horizon $H(t_b)=0$.
	
	\item[Growing Horizon ($0 < t < t_c$)]  
	As time progresses:
	\begin{itemize}
		\item Matter contracts: $h < R(t_2) < R(t_1) < 2h$ at sample times $t_1=0.3t_c$ (orange dotted) and $t_2=0.5t_c$ (cyan dotted)
		\item Horizon expands: $0 < H(t_1) < H(t_2) < h$ with intersections marked by orange/cyan stars
	\end{itemize}
	
	\item[Maximum Horizon ($t=t_c $)]  
	At critical time $t=t_c$, the matter boundary reaches $R(t_c)=h$ (blue solid line), coinciding with the maximal horizon size $H(t_c)=R(t_c)=h$ (blue star). For all $t\geq\,t_c$, the apparent horizon and event horizon become identical.
	
	\item[Constant Horizon ($t_c < t < t_d$)]  
	For $t_3=0.4(t_d-t_c)+t_c$ (magenta dotted line) and  $t_3=0.6(t_d-t_c)+t_c$ (purple dotted line) :
	\begin{itemize}
		\item Matter further collapses: $0< R(t_4) < R(t_3) < h$
		\item Horizon stabilizes: $H(t_3)=H(t_4)=h$ for $t_c < t_3 <  t_4 < t_d$
		\item {Resultant} spacetime structure:
		\begin{itemize}
			\item Schwarzschild vacuum for $r \geq R(t_{3,4})$
			\item Remaining matter in $0 < r < R(t_{3,4})$
		\end{itemize}
	\end{itemize}
	\item[Schwarzschild BH ($ t=t_d$)]  At the final time $t=t_d$ (red dashed line)  the collapsing star becomes the Schwarzschild BH.
\end{description}

{As already mentions, the EMT satisfies NEC, WEC and SEC for $2h < t_d < 4h$.} For $t < 0$, the horizon vanishes while the weak curvature singularity at $r=0$, {similar to} Eqs.~\eqref{scalars2} and \eqref{krets2}, becomes naked. To remedy this, we next modify the mass function to eliminate the naked singularity while preserving the energy conditions as much as possible.

\subsection{Removal of the naked singularity}
\label{remove}
To avoid the naked singularity at $r=0$  for $ t < 0$ before the formation of the BH, we modify the mass function as
\begin{eqnarray}
m(r,t) &=&\left\{
\begin{array}{ll}
h f\left( X(r,t) \right),  &X(r,t) \le 1,     \\
\\
 {\cal M}:= \dfrac{h}{2}, & X(r,t)  > 1 ,    \\
\end{array}
\right.
\label{mass3}
\end{eqnarray}
 where 
 \begin{eqnarray}
X(r,t) &=&\left\{
\begin{array}{ll}
x(r,t),  &t \ge 0,     \\
\\
x(r,t) \dfrac{r^\beta} {Q(t) +r^\beta}, & t < 0 .    \\
\end{array}
\right.
\label{mass33}
\end{eqnarray}

{The function $Q(t)$ satisfies the boundary conditions $Q(0)=\dot Q(0)=\ddot Q(0)=0$ while maintaining $Q(t) > 0$ for all $t < 0$. 
A natural smooth realization is given by the power series
\begin{eqnarray}
Q(t) = \sum_{n=3} a_n (-t)^n, \quad a_n \ge 0,
\end{eqnarray}
which ensures analytic behavior near $t=0$. More radically, one could consider}
\begin{eqnarray}
	\label{Q}
Q(t) = a_0 \exp\left[ \frac{c_0}{t}\right], \quad a_0 >0, c_0>0, 
\end{eqnarray}
whose $n$-th derivative vanishes at $t=0$ for all $n$, so that the EMT is smoothly connected at $t=0$. 
Therefore,  the expressions in Eqs.~\eqref{mass3} and~\eqref{mass33} remain perfectly regular,
even though $Q(t)$ in Eq.~\eqref{Q} has an essential singularity at $t=0$ ({\it i.e.} no analytic continuation is possible from $t<0$ to $t>0$).
In our numerical analysis, we adopt
\begin{equation}
Q(t) = {a_0 (-t)^3\over 1+ a_1(-t)^3}, \quad a_0>0, a_1>0,
\end{equation}
which satisfies $Q(t)\to a_0 (-t)^3$ at $t\to 0$, while $Q(t)\to a_0/a_1$ at $t\to-\infty$.

Note that  $a_n$ and $c_0$ are parameters necessary only for the negative $t$ but
none of such information is {required to describe} the BH at $t \ge 0$.

\subsubsection{Absence of singularities at $t<0$}
Derivatives of the mass function in Eqs.~\eqref{d_mass} and \eqref{dd_mass} can be easily generalized to the present case as
 \begin{eqnarray}
m_a(r,t) &=& h f^\prime \left( X(r,t) \right) X_a(r,t), 
\label{d_mass2} \\
m_{ab} (r,t) &=& h  f^\prime \left( X(r,t) \right) X_{ab}(r,t)\nonumber \\
 &+& h f^{\prime\prime} \left( X(r,t) \right) X_a(r,t) X_b(r,t) 
 \label{dd_mass2}
\end{eqnarray}

For a negative $t$ ($ t < 0$), 
$X$ and its derivatives near $r=0$ behaves as
\begin{eqnarray}
X, X_t, X_{tt} \sim r^{1+\beta}, \
X_r, X_{rt} \sim   r^{\beta}, \
X_{rr}\sim r^{\beta-1} .
\end{eqnarray}
Thus the most singular part of the EMT in Eqs.~\eqref{Ttt} $\sim$ \eqref{Ttr} near $r=0$ reads
\begin{eqnarray}
T^t{}_t, T^r{}_r, T^\theta{}_\theta=T^\phi{}_\phi \sim r^{\beta-2}, \quad
T^t{}_r, T^r{}_t \sim r^{\beta-1}.
\end{eqnarray}
Therefore, the singularity in the EMT (equivalently the Einstein tensor) at $r=0$ is absent for $\beta \ge 2$. 

\subsubsection{Continuity of the EMT at $X(r,t)=1$}
We next check continuity of the EMT  at $X(r,t)=1$ in both $r$ and $t$. Since the EMT vanishes at $X(r,t) > 1$, the EMT should also vanish at $X(r,t)=1$.

Denoting $r_s$ is a solution to $X(r_s,t)=1$ for $t< 0$, we see that
\begin{eqnarray}
m_a(r_s,t) &=& h X_a(r_s,t) f^\prime(1) =0, \\
m_{ab}(r_s,t) &=& h\left[ f^\prime(1) X_{ab}(r_s,t) + f^{\prime\prime}(1) X_{ab}(r_s,t)\right]=0,\nonumber\\
\end{eqnarray}
where we use properties of $f(x)$ that $f^\prime(1)=f^{\prime\prime}(1)=0$.
Therefore the EMT vanishes at $r=r_s$ and thus is continuous to the Schwarzschild vacuum at $ r> r_s$.

\subsubsection{Continuity of the EMT at $t=0$}
We also need to check continuity of the EMT at $t=0$. Using properties of $Q(t)$, we can explicitly show that
\begin{eqnarray}
X(r,0^-)&:=& \lim_{t\to 0^-} X(r,t) = x(r,0^+),\nonumber \\
X_r(r,0^-) &=& x_r(r,0)=x_r(r,0^+), \nonumber \\ 
 X_{rr}(r,0^-) &=& 0 =x_{rr}(r,0^+),\nonumber \\
X_{t}(r,0^-) &=&x_t(r,0)=x_t(r,0^+), \ \nonumber \\  
X_{tt}(r,0^-) &=&x_{tt}(r,0)=x_{tt}(r,0^+), \nonumber \\
X_{tr}(r,0^-) &=&x_{tr}(r,0)=x_{tr}(r,0^+).
\end{eqnarray}
The EMT in the $t\to 0^-$ limit agrees with that in the $t\to 0^+$ limit.
Therefore, the EMT is continuous at $t=0$.

\subsubsection{Energy conditions}
We investigate energy conditions for the EMT. While details will be given in the appendix \ref{app:EnergyCondition}, we here present only results without derivations.

A condition that $\beta\le 2$ is necessary for the NEC to be satisfied.  Combining the condition that $\beta \ge 2$ for the absence of the naked singularity,  we obtain $\beta=2$.

We have found that NEC and SEC are satisfied as long as we take $ 2h < t_d < 4h$.
Unfortunately, however, the SEC is always violated at $\beta > 0$.\\

Note that the total energy of matter is conserved during the whole process of the collapse
and agrees with the Schwarzschild BH mass (${\cal M}={h\over 2}$),
as in the case of the previous paper~\cite{Aoki:2024dyr}, 
since  we always have $T^t{}_t = -\frac{2 m_r}{\kappa r^2}$ for all $t$.
This result embodies (beyond Birkhoff's theorem) the complete incorporation of all collapsing matter within the final horizon without any external remnant.

\subsubsection{Mass function during a whole process of the gravitational collapse}
 \begin{figure}
	\centering
	\includegraphics[width=0.50\textwidth]{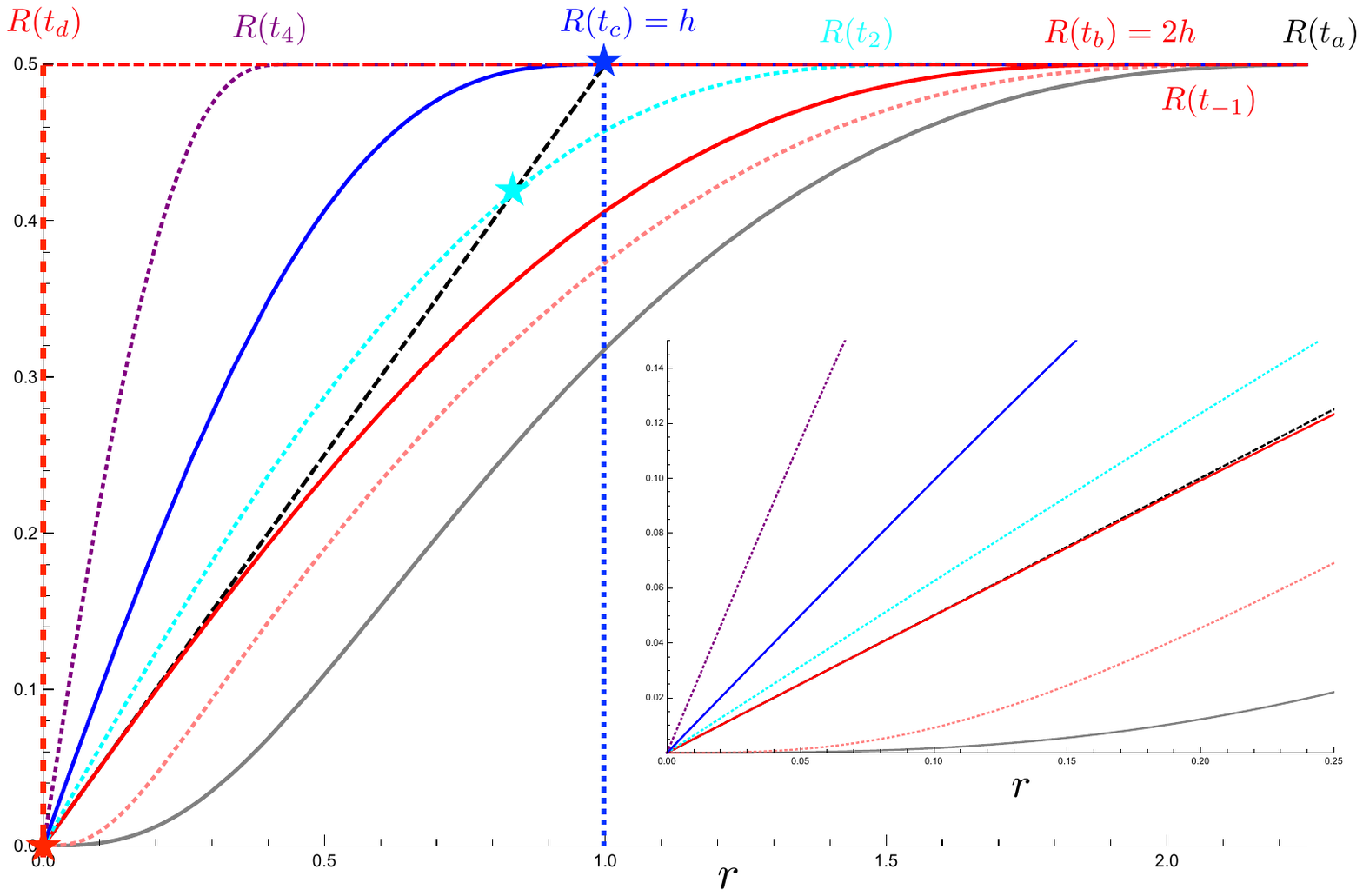}
	\caption{ A profile of $m(r,t)$ as a function of $r$ including negative $t$ such as $t_a= -3h/2$ (gray solid line), which is the starting time of the collapse,
	 and $t_{-1}= 0.5 t_a= -3h/4$ (pink dotted line).  The inset shows an enlargement near $r=0$.  See the main text for detailed explanations.  }
	\label{Fig4}
\end{figure}
 \begin{figure}
	\centering
	\includegraphics[width=0.33\textwidth]{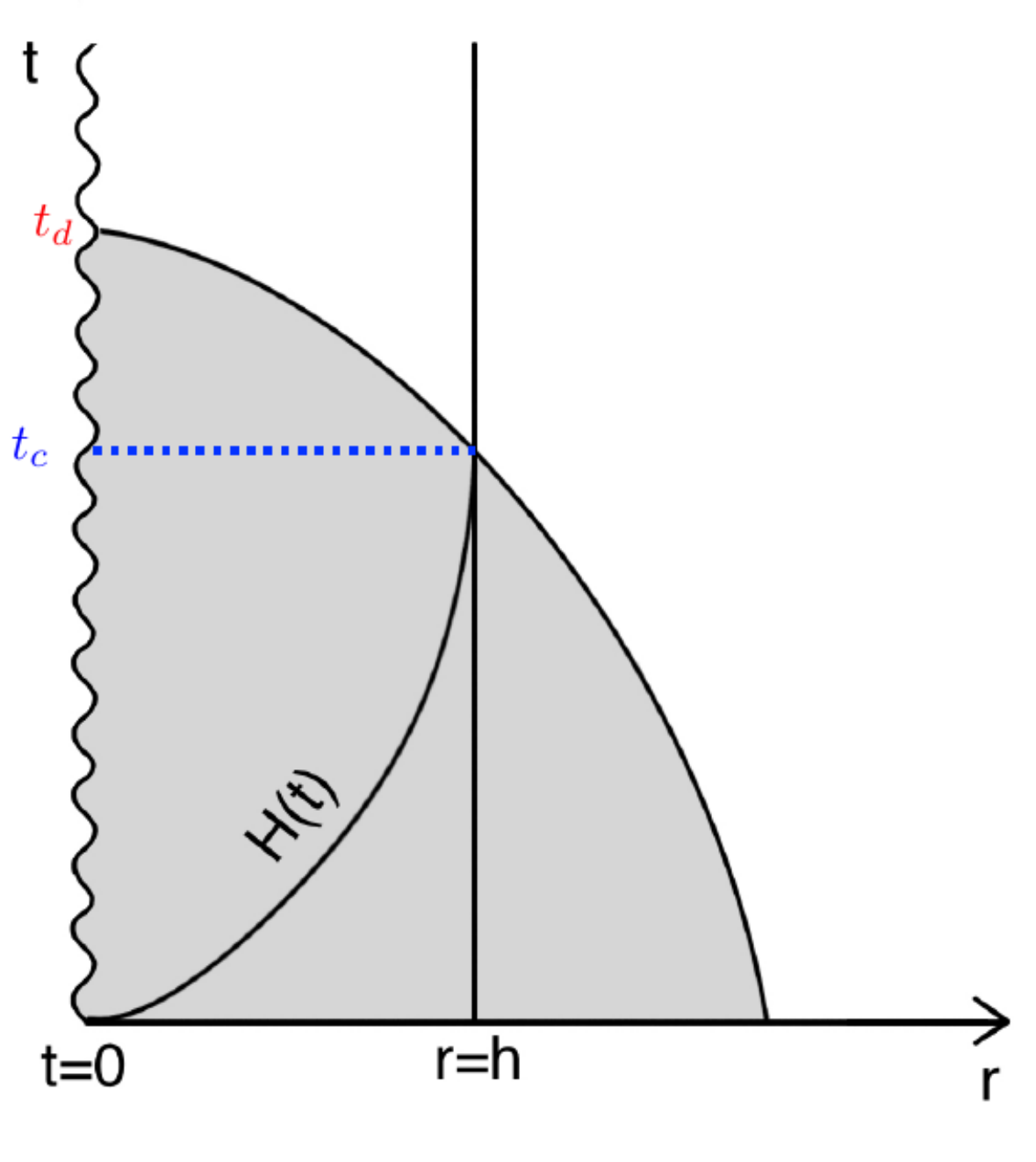}\\
		\includegraphics[width=0.38\textwidth]{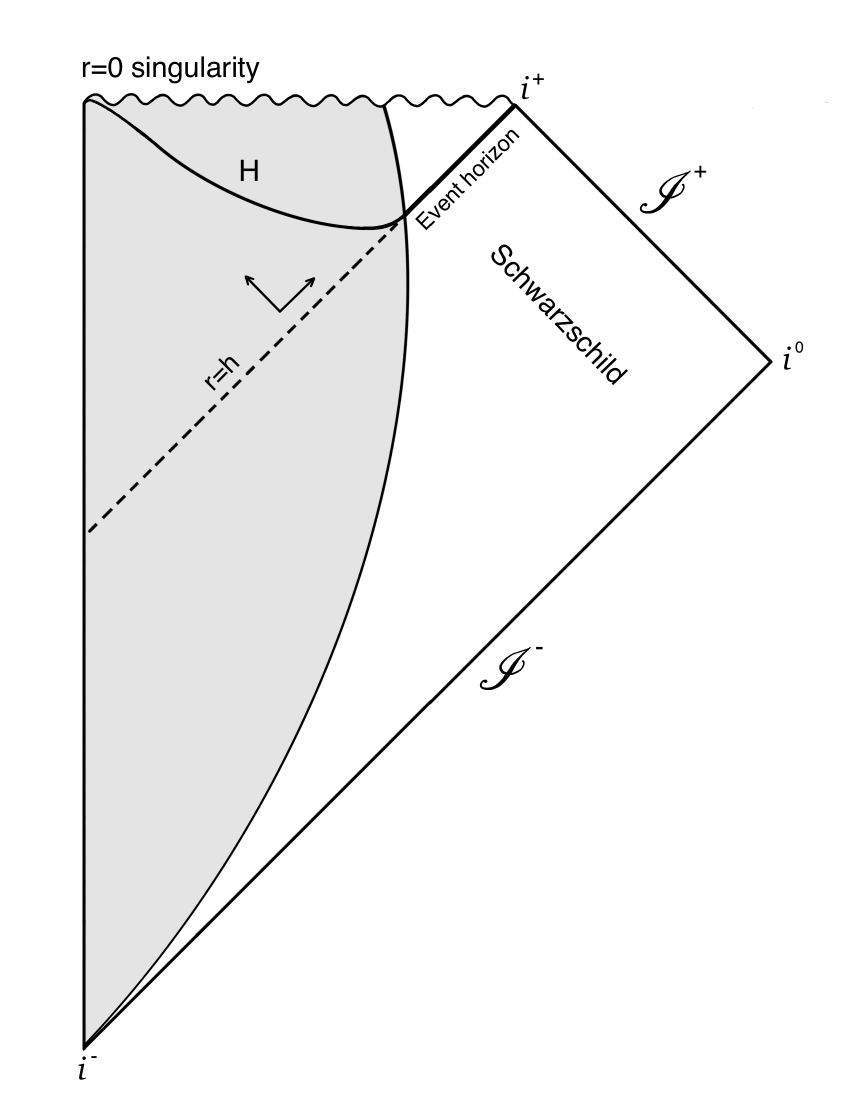}
	\caption{Representation of Gravitational Collapse. The shaded region depicts collapsing matter.
		Top: Spacetime diagram. The intersection of the hypersurface $r=h=2\mathcal{M}$ with the vacuum exterior represents the event horizon, while $H(t)$  marks the time-evolving apparent horizon.	
		Bottom: Corresponding Penrose (conformal) diagram, showing the formation of a spacelike central singularity.}
	\label{Figx}
\end{figure}
Fig.~\ref{Fig4} shows $m(r,t)$ at $\beta=2$ as a function of $r$ for several values of $t$ during a whole process of the collapse at $ t_a \le t \le t_d$.
where the simplest function $f(x)=f_0(x)$ is adopted and parameters are taken to be $h=1$, $t_d= 3h$ and $a_0=a_1=h/10$.
In the figure we add two  values of negative $t$  such as $t=t_a=-3h/2$ (gray) and $t_{-1}= 0.5 t_a=-3h/4$ (pink) in addition to 5 values ($t_b ,t_2, t_c,t_4,t_d$ except $t_1$ and $t_3$ for visibility) in Fig.~\ref{Fig3}.  As $t_a$ is the initial time of the collapse, $R(t_a)= 9h/4$ is {roughly} the largest radius of the matter distribution during the history of the collapse.
As can bee seen in the inset, $m(r,t) \sim r^3$ near $r=0$ for $t_a$ (gray) and $t_{-1}$ (orange) (negative $t$), avoiding the naked singularity at $r=0$, while $m(r,t) \sim r$ for positive $t$, producing the singularity at $r=0$.\\

Let us summarize a whole  history of the gravitational collapse bellow.

The gravitational collapse proceeds through distinct evolutionary phases. The collapse starts at  $t = t_{a} = -3h/2$, the matter distribution (gray solid line) extends to {around}  $R(t_a) = 9h/4 > 2h$ with no horizon present, as the mass profile $m(r,t_{a})$ never satisfies the horizon condition $m = r/2$ (black dashed line). Crucially, the cubic behavior $m(r,t_{a}) \sim r^3$ near the origin prevents naked singularity formation.

The horizon starts forming at $t = t_b = 0$ when $R(t_b) = 2h$ (red solid line), appearing at $r = 0$ (red star) as the slope condition $m_r(0,t_b) = 1/2$ is marginally satisfied. This marks both horizon formation with $H(t_b) = 0$ and the onset of central singularity development.

During $t_b=0 < t < t_c$, the horizon grows continuously while the matter contracts. At intermediate time $t = t_2$, the horizon reaches $0 < H(t_2) < h$ (cyan star), with the cyan dotted line maintaining $m_r(0,t_2) > 1/2$. The horizon reaches to its maximum size  as $H(t_c)=h=R(t_c)$ at $t=t_c$ (the blue solid line), shown by {the blue star as an} intersection with the horizon condition.

For $t_c < t_4 < t_d$, the matter boundary $R(t_4)$ contracts below $h$ (purple dotted line), creating a vacuum region $R(t_4) < r < h$ while the horizon remains fixed at $H(t_4) = h$. The system becomes a Schwarzschild BH at $t=t_d$.

We conclude emphasizing some key features of this model: (i) Monotonic horizon growth from zero to final size $h$, (ii) {Conserved energy with NEC and WEC fulfilled}
 throughout collapse, (iii) Avoidance of naked singularities and (iv) Smooth transition from matter distribution to vacuum solution. Fig.~\ref{Figx} summarizes the main features of the model.
 
 The singularity located at $r=0$ in the Penrose diagram is space-like, except at the onset of its formation ($t=0$, upper-left corner), where the hypersurface $r=0$ is null. The apparent horizon is likewise space-like for $0 \le t < t_c$, becoming null at $t=t_c$ and smoothly joining the event horizon.
Since the metric remains finite even at the singularity, there is no indication that the region at $r=0$ and $t=0$ extends along the null direction, in contrast to the behavior found in the Vaidya spacetime~\cite{Wheeler:2022xdo}. Rather, the available evidence suggests that this null character is confined to the event marking the onset of the singularity. Establishing this rigorously would require an analysis of null geodesics analogous to that performed in Ref.~\cite{Wheeler:2022xdo}. However, unlike the Vaidya case, the present geometry does not admit a comparably tractable analytic treatment, preventing a definitive conclusion from being obtained by the same method

\section{Conclusion}
\label{con}

\noindent 
In Ref.~\cite{Aoki:2024dyr}, we developed exact analytic models describing gravitational collapse from BHs with integrable singularities~\cite{Ovalle:2024wtv} to the Schwarzschild solution, maintaining a constant horizon size $h = 2\mathcal{M}$ throughout the process, where $\mathcal{M}$ is the BH mass. The present work extends this framework to earlier stages of collapse, while also resolving the causality issue present in Model II of Ref.~\cite{Aoki:2024dyr}, thereby providing a complete description of matter collapse from horizonless stellar configurations to BH formation. Our model precisely tracks the BH's evolution from its initial microscopic state ($h  \simeq  0$) to the final macroscopic configuration, offering new insights into the full dynamical process of gravitational collapse.

The exact analytical description of BH formation presented here, in particular the time-dependent evolution of the (apparent) horizon $H(t)$, offers insight into two features of gravitational collapse within the context of our model: (1) the dynamical emergence of trapped surfaces during gravitational collapse, and (2) the robust enforcement of cosmic censorship through controlled singularity formation. Our results illustrate how, in this setting, spacetime singularities become causally separated from distant observers via horizon growth (See Ref.~\cite{Bengtsson:2013hla} for a detailed discussion of the Oppenheimer-Snyder model), while offering a possible framework to study the quantum-to-classical transition in BH birth, i.e., scale bridging, connecting quantum gravitational regimes ($H(t) \sim \ell_{\rm Planck}$) to classical BH physics through the parameter $t_d$, which governs the collapse timescale.

\subsection*{Acknowledgments}
\vspace*{1mm}

{The authors would like to thank Prof. Jos\'{e} Senovilla for his useful comments.}
JO acknowledges financial support from the International Research Unit of Quantum
Information (QIU) of Kyoto University Research Coordination Alliance, 
and kind hospitality during his stay at Yukawa Institute for Theoretical Physics, Kyoto University. 
This work is partially supported by ANID FONDECYT Grant No. 1250227.
SA is supported in part by  the Grant-in-Aid of the Japanese Ministry of Education, Sports, Culture, Sciences and Technology (MEXT) for Scientific Research (No.~JP22H00129). 
\appendix

\section{{Energy conditions}}
\label{app:EnergyCondition}
In the previous paper~\cite{Aoki:2024dyr}, 
we have investigated energy conditions for the EMT given in  Eqs.~\eqref{Ttt} $\sim$ \eqref{Ttr}, whose form 
corresponds to the Type II  EMT of the Hawking-Ellis (Segr\'e-Pleba\'nski) classification~\cite{Plebanski:1964,Hawking:1973uf,Martin-Moruno:2017exc}.
Conditions in terms of $A,B,C$ in Eqs.~\eqref{Ttt} $\sim$ \eqref{Ttr},
which agree with those from a general method in Refs.~\cite{Maeda:2018hqu,Maeda:2022vld},
are summarized below (see also Refs.~\cite{Wang:2026jvo,Wang:2026sqr} for recent related developments).

\begin{description}
\item[ NEC]  The null energy condition is given as
\begin{eqnarray}
 A-B-C \ge 0.
\label{eq:NEC}
\end{eqnarray}

 \item[WEC] The weak energy condition leads to
\begin{eqnarray}
A-B-C \ge 0, \qquad A-B \ge 0.
\label{eq:WEC}
\end{eqnarray}

\item[SEC] The strong energy condition reads
\begin{eqnarray}
A-B-C \ge 0, \qquad C \le 0.
\label{eq:SEC}
\end{eqnarray}

\end{description}

We investigate various energy conditions  for $ t_a\le t \le t_d$ , using the new $R(t)$ in eq.~\eqref{eq:simpleR}.

\subsection{Null Energy Condition (NEC)}
The NEC requires 
\begin{eqnarray}
&Y_{\rm NEC } &(r,t) :=\frac{\kappa r^3(A - B - C)}{h X(1-X)} \nonumber \\
&=&d_1\frac{\dot R^2 r^2}{R^2}+ d_2 \frac{\ddot R r^2}{R} + 2d_3 \frac{\dot R r}{R} + d_4\ge 0,
\end{eqnarray}
where 
\begin{eqnarray}
d_1 &=&  2( 5 X^2-X-1), \ d_2=1+X-2X^2,\\
d_3 &=& = 6 X^2 +(8X^2-X-1)\left({\beta Q\over g} +{\dot Q r\over g}\right),\\
d_4 &=&  2(1+X+X^2) +\{(1+\beta)(1+X)+2(5-\beta) X^2\} {\beta Q\over g}\nonumber \\
&+& 2( 5X^2-X-1) \left({\beta Q\over g} +{\dot Q r\over g}\right)^2 
+(1-X)(2X+1){\ddot Q r^2\over g} \nonumber \\
&+& 2\{\beta(1+X)+(6-2\beta) X^2\} {\dot Q r\over g}
\end{eqnarray}
with $g:= Q(t) + r^\beta$. Eq.~\eqref{eq:simpleR} gives $2R \ddot R =\dot R^2 -1$.  \\

Setting $ Q=\dot Q =\ddot Q=0$ for $ 0\le t \le t_d$, we write
\begin{eqnarray}
F(x,\dot R):=2Y_{\rm NEC }(r,t) &=& \tilde d_1 \dot R^2 + 2 \tilde d_2 \dot R + \tilde d_3,
\end{eqnarray}
where
\begin{eqnarray}
\tilde d_1(x)&=& 3x^2(2x-1)(3x+1), \ \tilde d_2(x)=12x^3, \nonumber \\
\tilde d_3(x) &=&2x^4-x^3+3x^2+4x+4.
\end{eqnarray}
Since $\tilde d_1(x)>0$ and $\tilde d_3(x) > 0$, it is easy to see that 
\begin{eqnarray}
\min_{-1\le \dot R\le 0} F(x,\dot R) & =&  4(1-x)^2(5x^2+3x+1) \ge 0~~~~ 
\end{eqnarray}
for $0\le x\le 1$. The NEC is satisfied for $ 0\le t \le t_d$.\\

The analysis for the negative $t$ ( $t_a\le t < 0$) is more involved.

In the $X\to 0$ (equivalently $r\to 0$) limit,  we have
\begin{eqnarray}
Y_{NEC} = -(\beta -2)(\beta+1) -2\beta\left({\dot Q\over Q} +{\dot R\over R}\right) r \ge 0,~~~~~
\end{eqnarray}
 which requires $ \beta \le 2$. Combining with the condition $\beta \ge 2$ for the absence of naked singularities, we conclude $\beta=2$.
 The next leading order term in $r $ is positive, since $\dot R/R \le  0$ and $\dot Q/Q < 0$. 
 Therefore the NEC is satisfied near $X=0$.

The opposite limit that $X\to 1$ leads to
\begin{eqnarray}
Y_{NEC} =  6\left( \frac{\dot R}{R} + 1 +\frac{\beta Q}{g}+\frac{\dot Q Qr}{g}\right)^2 \ge 0,
\end{eqnarray}
so that the NEC is satisfied near $X=1$.

For  $0 < X(r,t) < 1$, an explicit evaluation by Mathematica is required, and we observe that the NEC is always satisfied at $t_a \le t  < 0$ for $2h < t_d < 4h$.
Near $t_d=2h$, however, we need to take $a_0=a_1:=a$ in $Q$ small enough to satisfy the NEC.
For example, $a=(t_d-2h) h/10$ works well at $t_d=2.1 h$.\\ 

In conclusion, the NEC is satisfied at  $t_a \le t \le t_d$ for  $2 h < t_d < 4h$, as long as we properly take $a_0$ and $a_1$ in $Q$. 

\subsection{Weak Energy Condition (WEC)}
In addition to $Y_{\rm NEC}(r,t) \ge 0$, the WEC requires 
\begin{eqnarray}
Y_{\rm WEC} (r,t)&:=& \frac{\kappa r^3 (A-B)} {2 h X f^\prime(X)} \nonumber \\
&=& 
 1 +{2 Q\over g} +{\dot Q r\over g}  +\frac{\dot R r}{R} \ge 0,
\end{eqnarray}
which is satisfied at $ 0\le t \le t_d$ since $1+\dot R x \ge 1+\dot R = 2 b_1 T \ge 0$.
Furthermore, the above condition is satisfied as $r\to 0$.

An explicit  calculation by Mathematica for $ 0 < X(r,t) \le 1$  tells us that
the WEC is always satisfied at $t_a \le t \le t_d$ for $2h < t_d < 4h$. 

\subsection{Strong Energy Condition (SEC)}
In addition to the NEC, the SEC requires
\begin{eqnarray}
Y_{\rm SEC}:=\frac{ -\kappa r^2 R C}{2 h x(1-x)} &=& \tilde c_1 \dot R^2 + 2 \tilde c_2 \dot R + \tilde c_3 \ge 0~~
\end{eqnarray}
for $0\le t \le t_d$, 
where
\begin{eqnarray}
\tilde c_1 &=&  3x (2x-1)(3x+1), \
\tilde c_2 = 2(8x^2-x-1), \nonumber \\
\tilde c_3 &=& x(2x^2-x+11).
\end{eqnarray}
Since $Y_{\rm SEC}$ takes the minimum value at $\dot R=1$, we show 
\begin{eqnarray}
\min_{-1\le \dot R \le 0} Y_{\rm SEC} = 4(1-x)^2(5x+1) \ge 0.
\end{eqnarray}
The SEC is satisfied at $0\le t \le t_d$.

At the negative $t$ ($t_a\le t <0$), however, we have  
\begin{eqnarray}
\lim_{X\to 0} \frac{-\kappa r^4 C}{h X(1-X)}= -\beta(\beta+1) =-6,
\end{eqnarray}
so that  the SEC is violated in this limit.\\

In conclusion, the SEC is only satisfied at $ 0\le t\le t_d$, but violated at  $t_a\le t <0$.
 
%
%
%
\bibliography{references.bib}

\begin{thebibliography}{53}%
\makeatletter
\providecommand \@ifxundefined [1]{%
 \@ifx{#1\undefined}
}%
\providecommand \@ifnum [1]{%
 \ifnum #1\expandafter \@firstoftwo
 \else \expandafter \@secondoftwo
 \fi
}%
\providecommand \@ifx [1]{%
 \ifx #1\expandafter \@firstoftwo
 \else \expandafter \@secondoftwo
 \fi
}%
\providecommand \natexlab [1]{#1}%
\providecommand \enquote  [1]{``#1''}%
\providecommand \bibnamefont  [1]{#1}%
\providecommand \bibfnamefont [1]{#1}%
\providecommand \citenamefont [1]{#1}%
\providecommand \href@noop [0]{\@secondoftwo}%
\providecommand \href [0]{\begingroup \@sanitize@url \@href}%
\providecommand \@href[1]{\@@startlink{#1}\@@href}%
\providecommand \@@href[1]{\endgroup#1\@@endlink}%
\providecommand \@sanitize@url [0]{\catcode `\\12\catcode `\$12\catcode
  `\&12\catcode `\#12\catcode `\^12\catcode `\_12\catcode `\%12\relax}%
\providecommand \@@startlink[1]{}%
\providecommand \@@endlink[0]{}%
\providecommand \url  [0]{\begingroup\@sanitize@url \@url }%
\providecommand \@url [1]{\endgroup\@href {#1}{\urlprefix }}%
\providecommand \urlprefix  [0]{URL }%
\providecommand \Eprint [0]{\href }%
\providecommand \doibase [0]{http://dx.doi.org/}%
\providecommand \selectlanguage [0]{\@gobble}%
\providecommand \bibinfo  [0]{\@secondoftwo}%
\providecommand \bibfield  [0]{\@secondoftwo}%
\providecommand \translation [1]{[#1]}%
\providecommand \BibitemOpen [0]{}%
\providecommand \bibitemStop [0]{}%
\providecommand \bibitemNoStop [0]{.\EOS\space}%
\providecommand \EOS [0]{\spacefactor3000\relax}%
\providecommand \BibitemShut  [1]{\csname bibitem#1\endcsname}%
\let\auto@bib@innerbib\@empty
\bibitem [{\citenamefont {Oppenheimer}\ and\ \citenamefont
  {Snyder}(1939)}]{Oppenheimer:1939ue}%
  \BibitemOpen
  \bibfield  {author} {\bibinfo {author} {\bibfnamefont {J.~R.}\ \bibnamefont
  {Oppenheimer}}\ and\ \bibinfo {author} {\bibfnamefont {H.}~\bibnamefont
  {Snyder}},\ }\href {\doibase 10.1103/PhysRev.56.455} {\bibfield  {journal}
  {\bibinfo  {journal} {Phys. Rev.}\ }\textbf {\bibinfo {volume} {56}},\
  \bibinfo {pages} {455} (\bibinfo {year} {1939})}\BibitemShut {NoStop}%
\bibitem [{\citenamefont {Bondi}(1947)}]{Bondi:1947fta}%
  \BibitemOpen
  \bibfield  {author} {\bibinfo {author} {\bibfnamefont {H.}~\bibnamefont
  {Bondi}},\ }\href {\doibase 10.1093/mnras/107.5-6.410} {\bibfield  {journal}
  {\bibinfo  {journal} {Mon. Not. Roy. Astron. Soc.}\ }\textbf {\bibinfo
  {volume} {107}},\ \bibinfo {pages} {410} (\bibinfo {year}
  {1947})}\BibitemShut {NoStop}%
\bibitem [{\citenamefont {Christodoulou}(1984)}]{Christodoulou:1984mz}%
  \BibitemOpen
  \bibfield  {author} {\bibinfo {author} {\bibfnamefont {D.}~\bibnamefont
  {Christodoulou}},\ }\href {\doibase 10.1007/BF01223743} {\bibfield  {journal}
  {\bibinfo  {journal} {Commun. Math. Phys.}\ }\textbf {\bibinfo {volume}
  {93}},\ \bibinfo {pages} {171} (\bibinfo {year} {1984})}\BibitemShut
  {NoStop}%
\bibitem [{\citenamefont {Joshi}\ and\ \citenamefont
  {Dwivedi}(1993)}]{Joshi:1993zg}%
  \BibitemOpen
  \bibfield  {author} {\bibinfo {author} {\bibfnamefont {P.~S.}\ \bibnamefont
  {Joshi}}\ and\ \bibinfo {author} {\bibfnamefont {I.~H.}\ \bibnamefont
  {Dwivedi}},\ }\href {\doibase 10.1103/PhysRevD.47.5357} {\bibfield  {journal}
  {\bibinfo  {journal} {Phys. Rev. D}\ }\textbf {\bibinfo {volume} {47}},\
  \bibinfo {pages} {5357} (\bibinfo {year} {1993})},\ \Eprint
  {http://arxiv.org/abs/gr-qc/9303037} {arXiv:gr-qc/9303037} \BibitemShut
  {NoStop}%
\bibitem [{\citenamefont {Shibata}(1999)}]{Shibata:1999va}%
  \BibitemOpen
  \bibfield  {author} {\bibinfo {author} {\bibfnamefont {M.}~\bibnamefont
  {Shibata}},\ }\href {\doibase 10.1143/PTP.101.251} {\bibfield  {journal}
  {\bibinfo  {journal} {Prog. Theor. Phys.}\ }\textbf {\bibinfo {volume}
  {101}},\ \bibinfo {pages} {251} (\bibinfo {year} {1999})}\BibitemShut
  {NoStop}%
\bibitem [{\citenamefont {Joshi}\ \emph {et~al.}(2002)\citenamefont {Joshi},
  \citenamefont {Dadhich},\ and\ \citenamefont {Maartens}}]{Joshi:2001xi}%
  \BibitemOpen
  \bibfield  {author} {\bibinfo {author} {\bibfnamefont {P.~S.}\ \bibnamefont
  {Joshi}}, \bibinfo {author} {\bibfnamefont {N.}~\bibnamefont {Dadhich}}, \
  and\ \bibinfo {author} {\bibfnamefont {R.}~\bibnamefont {Maartens}},\ }\href
  {\doibase 10.1103/PhysRevD.65.101501} {\bibfield  {journal} {\bibinfo
  {journal} {Phys. Rev. D}\ }\textbf {\bibinfo {volume} {65}},\ \bibinfo
  {pages} {101501} (\bibinfo {year} {2002})},\ \Eprint
  {http://arxiv.org/abs/gr-qc/0109051} {arXiv:gr-qc/0109051} \BibitemShut
  {NoStop}%
\bibitem [{\citenamefont {Giambo}\ \emph {et~al.}(2004)\citenamefont {Giambo},
  \citenamefont {Giannoni}, \citenamefont {Magli},\ and\ \citenamefont
  {Piccione}}]{Giambo:2003fd}%
  \BibitemOpen
  \bibfield  {author} {\bibinfo {author} {\bibfnamefont {R.}~\bibnamefont
  {Giambo}}, \bibinfo {author} {\bibfnamefont {F.}~\bibnamefont {Giannoni}},
  \bibinfo {author} {\bibfnamefont {G.}~\bibnamefont {Magli}}, \ and\ \bibinfo
  {author} {\bibfnamefont {P.}~\bibnamefont {Piccione}},\ }\href {\doibase
  10.1023/B:GERG.0000022388.11306.e1} {\bibfield  {journal} {\bibinfo
  {journal} {Gen. Rel. Grav.}\ }\textbf {\bibinfo {volume} {36}},\ \bibinfo
  {pages} {1279} (\bibinfo {year} {2004})},\ \Eprint
  {http://arxiv.org/abs/gr-qc/0303043} {arXiv:gr-qc/0303043} \BibitemShut
  {NoStop}%
\bibitem [{\citenamefont {Goswami}\ and\ \citenamefont
  {Joshi}(2004)}]{Goswami:2003bs}%
  \BibitemOpen
  \bibfield  {author} {\bibinfo {author} {\bibfnamefont {R.}~\bibnamefont
  {Goswami}}\ and\ \bibinfo {author} {\bibfnamefont {P.~S.}\ \bibnamefont
  {Joshi}},\ }\href {\doibase 10.1103/PhysRevD.69.027502} {\bibfield  {journal}
  {\bibinfo  {journal} {Phys. Rev. D}\ }\textbf {\bibinfo {volume} {69}},\
  \bibinfo {pages} {027502} (\bibinfo {year} {2004})},\ \Eprint
  {http://arxiv.org/abs/gr-qc/0310122} {arXiv:gr-qc/0310122} \BibitemShut
  {NoStop}%
\bibitem [{\citenamefont {Lasky}\ and\ \citenamefont
  {Lun}(2006)}]{Lasky:2006mg}%
  \BibitemOpen
  \bibfield  {author} {\bibinfo {author} {\bibfnamefont {P.~D.}\ \bibnamefont
  {Lasky}}\ and\ \bibinfo {author} {\bibfnamefont {A.~W.~C.}\ \bibnamefont
  {Lun}},\ }\href {\doibase 10.1103/PhysRevD.74.084013} {\bibfield  {journal}
  {\bibinfo  {journal} {Phys. Rev. D}\ }\textbf {\bibinfo {volume} {74}},\
  \bibinfo {pages} {084013} (\bibinfo {year} {2006})},\ \Eprint
  {http://arxiv.org/abs/gr-qc/0606055} {arXiv:gr-qc/0606055} \BibitemShut
  {NoStop}%
\bibitem [{\citenamefont {Mosani}\ \emph {et~al.}(2020)\citenamefont {Mosani},
  \citenamefont {Dey},\ and\ \citenamefont {Joshi}}]{Mosani:2020ena}%
  \BibitemOpen
  \bibfield  {author} {\bibinfo {author} {\bibfnamefont {K.}~\bibnamefont
  {Mosani}}, \bibinfo {author} {\bibfnamefont {D.}~\bibnamefont {Dey}}, \ and\
  \bibinfo {author} {\bibfnamefont {P.~S.}\ \bibnamefont {Joshi}},\ }\href
  {\doibase 10.1103/PhysRevD.101.044052} {\bibfield  {journal} {\bibinfo
  {journal} {Phys. Rev. D}\ }\textbf {\bibinfo {volume} {101}},\ \bibinfo
  {pages} {044052} (\bibinfo {year} {2020})},\ \bibinfo {note} {[Erratum:
  Phys.Rev.D 107, 069903 (2023)]},\ \Eprint {http://arxiv.org/abs/2001.04367}
  {arXiv:2001.04367 [gr-qc]} \BibitemShut {NoStop}%
\bibitem [{\citenamefont {Ben~Achour}\ \emph {et~al.}(2020)\citenamefont
  {Ben~Achour}, \citenamefont {Brahma}, \citenamefont {Mukohyama},\ and\
  \citenamefont {Uzan}}]{BenAchour:2020gon}%
  \BibitemOpen
  \bibfield  {author} {\bibinfo {author} {\bibfnamefont {J.}~\bibnamefont
  {Ben~Achour}}, \bibinfo {author} {\bibfnamefont {S.}~\bibnamefont {Brahma}},
  \bibinfo {author} {\bibfnamefont {S.}~\bibnamefont {Mukohyama}}, \ and\
  \bibinfo {author} {\bibfnamefont {J.~P.}\ \bibnamefont {Uzan}},\ }\href
  {\doibase 10.1088/1475-7516/2020/09/020} {\bibfield  {journal} {\bibinfo
  {journal} {JCAP}\ }\textbf {\bibinfo {volume} {09}},\ \bibinfo {pages} {020}
  (\bibinfo {year} {2020})},\ \Eprint {http://arxiv.org/abs/2004.12977}
  {arXiv:2004.12977 [gr-qc]} \BibitemShut {NoStop}%
\bibitem [{\citenamefont {Joshi}\ and\ \citenamefont
  {Malafarina}(2011)}]{Joshi:2011rlc}%
  \BibitemOpen
  \bibfield  {author} {\bibinfo {author} {\bibfnamefont {P.~S.}\ \bibnamefont
  {Joshi}}\ and\ \bibinfo {author} {\bibfnamefont {D.}~\bibnamefont
  {Malafarina}},\ }\href {\doibase 10.1142/S0218271811020792} {\bibfield
  {journal} {\bibinfo  {journal} {Int. J. Mod. Phys. D}\ }\textbf {\bibinfo
  {volume} {20}},\ \bibinfo {pages} {2641} (\bibinfo {year} {2011})},\ \Eprint
  {http://arxiv.org/abs/1201.3660} {arXiv:1201.3660 [gr-qc]} \BibitemShut
  {NoStop}%
\bibitem [{\citenamefont {Greenwood}\ and\ \citenamefont
  {Stojkovic}(2008)}]{Greenwood:2008ht}%
  \BibitemOpen
  \bibfield  {author} {\bibinfo {author} {\bibfnamefont {E.}~\bibnamefont
  {Greenwood}}\ and\ \bibinfo {author} {\bibfnamefont {D.}~\bibnamefont
  {Stojkovic}},\ }\href {\doibase 10.1088/1126-6708/2008/06/042} {\bibfield
  {journal} {\bibinfo  {journal} {JHEP}\ }\textbf {\bibinfo {volume} {06}},\
  \bibinfo {pages} {042} (\bibinfo {year} {2008})},\ \Eprint
  {http://arxiv.org/abs/0802.4087} {arXiv:0802.4087 [gr-qc]} \BibitemShut
  {NoStop}%
\bibitem [{\citenamefont {Wang}\ \emph {et~al.}(2009)\citenamefont {Wang},
  \citenamefont {Greenwood},\ and\ \citenamefont {Stojkovic}}]{Wang:2009ay}%
  \BibitemOpen
  \bibfield  {author} {\bibinfo {author} {\bibfnamefont {J.~E.}\ \bibnamefont
  {Wang}}, \bibinfo {author} {\bibfnamefont {E.}~\bibnamefont {Greenwood}}, \
  and\ \bibinfo {author} {\bibfnamefont {D.}~\bibnamefont {Stojkovic}},\ }\href
  {\doibase 10.1103/PhysRevD.80.124027} {\bibfield  {journal} {\bibinfo
  {journal} {Phys. Rev. D}\ }\textbf {\bibinfo {volume} {80}},\ \bibinfo
  {pages} {124027} (\bibinfo {year} {2009})},\ \Eprint
  {http://arxiv.org/abs/0906.3250} {arXiv:0906.3250 [hep-th]} \BibitemShut
  {NoStop}%
\bibitem [{\citenamefont {Kiefer}\ and\ \citenamefont
  {Schmitz}(2019)}]{Kiefer:2019csi}%
  \BibitemOpen
  \bibfield  {author} {\bibinfo {author} {\bibfnamefont {C.}~\bibnamefont
  {Kiefer}}\ and\ \bibinfo {author} {\bibfnamefont {T.}~\bibnamefont
  {Schmitz}},\ }\href {\doibase 10.1103/PhysRevD.99.126010} {\bibfield
  {journal} {\bibinfo  {journal} {Phys. Rev. D}\ }\textbf {\bibinfo {volume}
  {99}},\ \bibinfo {pages} {126010} (\bibinfo {year} {2019})},\ \Eprint
  {http://arxiv.org/abs/1904.13220} {arXiv:1904.13220 [gr-qc]} \BibitemShut
  {NoStop}%
\bibitem [{\citenamefont {Blanchette}\ \emph {et~al.}(2021)\citenamefont
  {Blanchette}, \citenamefont {Das}, \citenamefont {Hergott},\ and\
  \citenamefont {Rastgoo}}]{Blanchette:2020kkk}%
  \BibitemOpen
  \bibfield  {author} {\bibinfo {author} {\bibfnamefont {K.}~\bibnamefont
  {Blanchette}}, \bibinfo {author} {\bibfnamefont {S.}~\bibnamefont {Das}},
  \bibinfo {author} {\bibfnamefont {S.}~\bibnamefont {Hergott}}, \ and\
  \bibinfo {author} {\bibfnamefont {S.}~\bibnamefont {Rastgoo}},\ }\href
  {\doibase 10.1103/PhysRevD.103.084038} {\bibfield  {journal} {\bibinfo
  {journal} {Phys. Rev. D}\ }\textbf {\bibinfo {volume} {103}},\ \bibinfo
  {pages} {084038} (\bibinfo {year} {2021})},\ \Eprint
  {http://arxiv.org/abs/2011.11815} {arXiv:2011.11815 [gr-qc]} \BibitemShut
  {NoStop}%
\bibitem [{\citenamefont {Piechocki}\ and\ \citenamefont
  {Schmitz}(2020)}]{Piechocki:2020bfo}%
  \BibitemOpen
  \bibfield  {author} {\bibinfo {author} {\bibfnamefont {W.}~\bibnamefont
  {Piechocki}}\ and\ \bibinfo {author} {\bibfnamefont {T.}~\bibnamefont
  {Schmitz}},\ }\href {\doibase 10.1103/PhysRevD.102.046004} {\bibfield
  {journal} {\bibinfo  {journal} {Phys. Rev. D}\ }\textbf {\bibinfo {volume}
  {102}},\ \bibinfo {pages} {046004} (\bibinfo {year} {2020})},\ \Eprint
  {http://arxiv.org/abs/2004.02939} {arXiv:2004.02939 [gr-qc]} \BibitemShut
  {NoStop}%
\bibitem [{\citenamefont {Kelly}\ \emph {et~al.}(2021)\citenamefont {Kelly},
  \citenamefont {Santacruz},\ and\ \citenamefont
  {Wilson-Ewing}}]{Kelly:2020lec}%
  \BibitemOpen
  \bibfield  {author} {\bibinfo {author} {\bibfnamefont {J.~G.}\ \bibnamefont
  {Kelly}}, \bibinfo {author} {\bibfnamefont {R.}~\bibnamefont {Santacruz}}, \
  and\ \bibinfo {author} {\bibfnamefont {E.}~\bibnamefont {Wilson-Ewing}},\
  }\href {\doibase 10.1088/1361-6382/abd3e2} {\bibfield  {journal} {\bibinfo
  {journal} {Class. Quant. Grav.}\ }\textbf {\bibinfo {volume} {38}},\ \bibinfo
  {pages} {04LT01} (\bibinfo {year} {2021})},\ \Eprint
  {http://arxiv.org/abs/2006.09325} {arXiv:2006.09325 [gr-qc]} \BibitemShut
  {NoStop}%
\bibitem [{\citenamefont {Husain}\ \emph
  {et~al.}(2022{\natexlab{a}})\citenamefont {Husain}, \citenamefont {Kelly},
  \citenamefont {Santacruz},\ and\ \citenamefont
  {Wilson-Ewing}}]{Husain:2021ojz}%
  \BibitemOpen
  \bibfield  {author} {\bibinfo {author} {\bibfnamefont {V.}~\bibnamefont
  {Husain}}, \bibinfo {author} {\bibfnamefont {J.~G.}\ \bibnamefont {Kelly}},
  \bibinfo {author} {\bibfnamefont {R.}~\bibnamefont {Santacruz}}, \ and\
  \bibinfo {author} {\bibfnamefont {E.}~\bibnamefont {Wilson-Ewing}},\ }\href
  {\doibase 10.1103/PhysRevLett.128.121301} {\bibfield  {journal} {\bibinfo
  {journal} {Phys. Rev. Lett.}\ }\textbf {\bibinfo {volume} {128}},\ \bibinfo
  {pages} {121301} (\bibinfo {year} {2022}{\natexlab{a}})},\ \Eprint
  {http://arxiv.org/abs/2109.08667} {arXiv:2109.08667 [gr-qc]} \BibitemShut
  {NoStop}%
\bibitem [{\citenamefont {Husain}\ \emph
  {et~al.}(2022{\natexlab{b}})\citenamefont {Husain}, \citenamefont {Kelly},
  \citenamefont {Santacruz},\ and\ \citenamefont
  {Wilson-Ewing}}]{Husain:2022gwp}%
  \BibitemOpen
  \bibfield  {author} {\bibinfo {author} {\bibfnamefont {V.}~\bibnamefont
  {Husain}}, \bibinfo {author} {\bibfnamefont {J.~G.}\ \bibnamefont {Kelly}},
  \bibinfo {author} {\bibfnamefont {R.}~\bibnamefont {Santacruz}}, \ and\
  \bibinfo {author} {\bibfnamefont {E.}~\bibnamefont {Wilson-Ewing}},\ }\href
  {\doibase 10.1103/PhysRevD.106.024014} {\bibfield  {journal} {\bibinfo
  {journal} {Phys. Rev. D}\ }\textbf {\bibinfo {volume} {106}},\ \bibinfo
  {pages} {024014} (\bibinfo {year} {2022}{\natexlab{b}})},\ \Eprint
  {http://arxiv.org/abs/2203.04238} {arXiv:2203.04238 [gr-qc]} \BibitemShut
  {NoStop}%
\bibitem [{\citenamefont {Casadio}\ \emph {et~al.}(2023)\citenamefont
  {Casadio}, \citenamefont {da~Rocha}, \citenamefont {Meert}, \citenamefont
  {Tabarroni},\ and\ \citenamefont {Barreto}}]{Casadio:2022pla}%
  \BibitemOpen
  \bibfield  {author} {\bibinfo {author} {\bibfnamefont {R.}~\bibnamefont
  {Casadio}}, \bibinfo {author} {\bibfnamefont {R.}~\bibnamefont {da~Rocha}},
  \bibinfo {author} {\bibfnamefont {P.}~\bibnamefont {Meert}}, \bibinfo
  {author} {\bibfnamefont {L.}~\bibnamefont {Tabarroni}}, \ and\ \bibinfo
  {author} {\bibfnamefont {W.}~\bibnamefont {Barreto}},\ }\href {\doibase
  10.1088/1361-6382/acbe89} {\bibfield  {journal} {\bibinfo  {journal} {Class.
  Quant. Grav.}\ }\textbf {\bibinfo {volume} {40}},\ \bibinfo {pages} {075014}
  (\bibinfo {year} {2023})},\ \Eprint {http://arxiv.org/abs/2206.10398}
  {arXiv:2206.10398 [gr-qc]} \BibitemShut {NoStop}%
\bibitem [{\citenamefont {Cipriani}\ \emph {et~al.}(2024)\citenamefont
  {Cipriani}, \citenamefont {Fazzini},\ and\ \citenamefont
  {Wilson-Ewing}}]{Cipriani:2024nhx}%
  \BibitemOpen
  \bibfield  {author} {\bibinfo {author} {\bibfnamefont {L.}~\bibnamefont
  {Cipriani}}, \bibinfo {author} {\bibfnamefont {F.}~\bibnamefont {Fazzini}}, \
  and\ \bibinfo {author} {\bibfnamefont {E.}~\bibnamefont {Wilson-Ewing}},\
  }\href {\doibase 10.1103/PhysRevD.110.066004} {\bibfield  {journal} {\bibinfo
   {journal} {Phys. Rev. D}\ }\textbf {\bibinfo {volume} {110}},\ \bibinfo
  {pages} {066004} (\bibinfo {year} {2024})},\ \Eprint
  {http://arxiv.org/abs/2404.04192} {arXiv:2404.04192 [gr-qc]} \BibitemShut
  {NoStop}%
\bibitem [{\citenamefont {Casadio}\ \emph {et~al.}(2026)\citenamefont
  {Casadio}, \citenamefont {Giusti}, \citenamefont {Kamenshchik},\ and\
  \citenamefont {Ovalle}}]{Casadio:2026tmd}%
  \BibitemOpen
  \bibfield  {author} {\bibinfo {author} {\bibfnamefont {R.}~\bibnamefont
  {Casadio}}, \bibinfo {author} {\bibfnamefont {A.}~\bibnamefont {Giusti}},
  \bibinfo {author} {\bibfnamefont {A.}~\bibnamefont {Kamenshchik}}, \ and\
  \bibinfo {author} {\bibfnamefont {J.}~\bibnamefont {Ovalle}},\ }\href@noop {}
  {\  (\bibinfo {year} {2026})},\ \Eprint {http://arxiv.org/abs/2605.01808}
  {arXiv:2605.01808 [gr-qc]} \BibitemShut {NoStop}%
\bibitem [{\citenamefont {Penrose}(1965)}]{Penrose:1964wq}%
  \BibitemOpen
  \bibfield  {author} {\bibinfo {author} {\bibfnamefont {R.}~\bibnamefont
  {Penrose}},\ }\href {\doibase 10.1103/PhysRevLett.14.57} {\bibfield
  {journal} {\bibinfo  {journal} {Phys. Rev. Lett.}\ }\textbf {\bibinfo
  {volume} {14}},\ \bibinfo {pages} {57} (\bibinfo {year} {1965})}\BibitemShut
  {NoStop}%
\bibitem [{\citenamefont {Akiyama}\ \emph {et~al.}(2019)\citenamefont {Akiyama}
  \emph {et~al.}}]{EventHorizonTelescope:2019dse}%
  \BibitemOpen
  \bibfield  {author} {\bibinfo {author} {\bibfnamefont {K.}~\bibnamefont
  {Akiyama}} \emph {et~al.} (\bibinfo {collaboration} {Event Horizon
  Telescope}),\ }\href {\doibase 10.3847/2041-8213/ab0ec7} {\bibfield
  {journal} {\bibinfo  {journal} {Astrophys. J. Lett.}\ }\textbf {\bibinfo
  {volume} {875}},\ \bibinfo {pages} {L1} (\bibinfo {year} {2019})},\ \Eprint
  {http://arxiv.org/abs/1906.11238} {arXiv:1906.11238 [astro-ph.GA]}
  \BibitemShut {NoStop}%
\bibitem [{\citenamefont {Akiyama}\ \emph {et~al.}(2022)\citenamefont {Akiyama}
  \emph {et~al.}}]{EventHorizonTelescope:2022wkp}%
  \BibitemOpen
  \bibfield  {author} {\bibinfo {author} {\bibfnamefont {K.}~\bibnamefont
  {Akiyama}} \emph {et~al.} (\bibinfo {collaboration} {Event Horizon
  Telescope}),\ }\href {\doibase 10.3847/2041-8213/ac6674} {\bibfield
  {journal} {\bibinfo  {journal} {Astrophys. J. Lett.}\ }\textbf {\bibinfo
  {volume} {930}},\ \bibinfo {pages} {L12} (\bibinfo {year} {2022})},\ \Eprint
  {http://arxiv.org/abs/2311.08680} {arXiv:2311.08680 [astro-ph.HE]}
  \BibitemShut {NoStop}%
\bibitem [{\citenamefont {Poisson}\ and\ \citenamefont
  {Israel}(1989)}]{Poisson:1989zz}%
  \BibitemOpen
  \bibfield  {author} {\bibinfo {author} {\bibfnamefont {E.}~\bibnamefont
  {Poisson}}\ and\ \bibinfo {author} {\bibfnamefont {W.}~\bibnamefont
  {Israel}},\ }\href {\doibase 10.1103/PhysRevLett.63.1663} {\bibfield
  {journal} {\bibinfo  {journal} {Phys. Rev. Lett.}\ }\textbf {\bibinfo
  {volume} {63}},\ \bibinfo {pages} {1663} (\bibinfo {year}
  {1989})}\BibitemShut {NoStop}%
\bibitem [{\citenamefont {Poisson}\ and\ \citenamefont
  {Israel}(1990)}]{Poisson:1990eh}%
  \BibitemOpen
  \bibfield  {author} {\bibinfo {author} {\bibfnamefont {E.}~\bibnamefont
  {Poisson}}\ and\ \bibinfo {author} {\bibfnamefont {W.}~\bibnamefont
  {Israel}},\ }\href {\doibase 10.1103/PhysRevD.41.1796} {\bibfield  {journal}
  {\bibinfo  {journal} {Phys. Rev. D}\ }\textbf {\bibinfo {volume} {41}},\
  \bibinfo {pages} {1796} (\bibinfo {year} {1990})}\BibitemShut {NoStop}%
\bibitem [{\citenamefont {Penrose}(1969)}]{Penrose:1969pc}%
  \BibitemOpen
  \bibfield  {author} {\bibinfo {author} {\bibfnamefont {R.}~\bibnamefont
  {Penrose}},\ }\href {\doibase 10.1023/A:1016578408204} {\bibfield  {journal}
  {\bibinfo  {journal} {Riv. Nuovo Cim.}\ }\textbf {\bibinfo {volume} {1}},\
  \bibinfo {pages} {252} (\bibinfo {year} {1969})}\BibitemShut {NoStop}%
\bibitem [{\citenamefont {Penrose}(1979)}]{Penrose1979}%
  \BibitemOpen
  \bibfield  {author} {\bibinfo {author} {\bibfnamefont {R.}~\bibnamefont
  {Penrose}},\ }in\ \href@noop {} {\emph {\bibinfo {booktitle} {General
  Relativity: An Einstein Centenary Survey}}},\ \bibinfo {editor} {edited by\
  \bibinfo {editor} {\bibfnamefont {S.~W.}\ \bibnamefont {Hawking}}\ and\
  \bibinfo {editor} {\bibfnamefont {W.}~\bibnamefont {Israel}}}\ (\bibinfo
  {publisher} {Cambridge University Press},\ \bibinfo {year} {1979})\ pp.\
  \bibinfo {pages} {581--638}\BibitemShut {NoStop}%
\bibitem [{\citenamefont {Wald}(1984)}]{Wald:1984rg}%
  \BibitemOpen
  \bibfield  {author} {\bibinfo {author} {\bibfnamefont {R.~M.}\ \bibnamefont
  {Wald}},\ }\href {\doibase 10.7208/chicago/9780226870373.001.0001} {\emph
  {\bibinfo {title} {{General Relativity}}}}\ (\bibinfo  {publisher} {Chicago
  Univ. Pr.},\ \bibinfo {address} {Chicago, USA},\ \bibinfo {year}
  {1984})\BibitemShut {NoStop}%
\bibitem [{\citenamefont {Ovalle}(2024)}]{Ovalle:2024wtv}%
  \BibitemOpen
  \bibfield  {author} {\bibinfo {author} {\bibfnamefont {J.}~\bibnamefont
  {Ovalle}},\ }\href {\doibase 10.1103/PhysRevD.109.104032} {\bibfield
  {journal} {\bibinfo  {journal} {Phys. Rev. D}\ }\textbf {\bibinfo {volume}
  {109}},\ \bibinfo {pages} {104032} (\bibinfo {year} {2024})}\BibitemShut
  {NoStop}%
\bibitem [{\citenamefont {Ovalle}(2025)}]{Ovalle:2025pue}%
  \BibitemOpen
  \bibfield  {author} {\bibinfo {author} {\bibfnamefont {J.}~\bibnamefont
  {Ovalle}},\ }\href@noop {} {\  (\bibinfo {year} {2025})},\ \Eprint
  {http://arxiv.org/abs/2509.00816} {arXiv:2509.00816 [gr-qc]} \BibitemShut
  {NoStop}%
\bibitem [{\citenamefont {Ovalle}\ \emph {et~al.}(2026)\citenamefont {Ovalle},
  \citenamefont {Casadio},\ and\ \citenamefont {Kamenshchik}}]{Ovalle:2026lxb}%
  \BibitemOpen
  \bibfield  {author} {\bibinfo {author} {\bibfnamefont {J.}~\bibnamefont
  {Ovalle}}, \bibinfo {author} {\bibfnamefont {R.}~\bibnamefont {Casadio}}, \
  and\ \bibinfo {author} {\bibfnamefont {A.}~\bibnamefont {Kamenshchik}},\
  }\href {\doibase 10.1103/cbs6-d7pr} {\bibfield  {journal} {\bibinfo
  {journal} {Phys. Rev. D}\ }\textbf {\bibinfo {volume} {113}},\ \bibinfo
  {pages} {064042} (\bibinfo {year} {2026})},\ \Eprint
  {http://arxiv.org/abs/2603.06451} {arXiv:2603.06451 [gr-qc]} \BibitemShut
  {NoStop}%
\bibitem [{\citenamefont {Aoki}\ and\ \citenamefont
  {Ovalle}(2025)}]{Aoki:2024dyr}%
  \BibitemOpen
  \bibfield  {author} {\bibinfo {author} {\bibfnamefont {S.}~\bibnamefont
  {Aoki}}\ and\ \bibinfo {author} {\bibfnamefont {J.}~\bibnamefont {Ovalle}},\
  }\href {\doibase 10.1103/PhysRevD.111.024037} {\bibfield  {journal} {\bibinfo
   {journal} {Phys. Rev. D}\ }\textbf {\bibinfo {volume} {111}},\ \bibinfo
  {pages} {024037} (\bibinfo {year} {2025})}\BibitemShut {NoStop}%
\bibitem [{\citenamefont {Casadio}\ \emph {et~al.}(2024)\citenamefont
  {Casadio}, \citenamefont {Kamenshchik},\ and\ \citenamefont
  {Ovalle}}]{Casadio:2024fol}%
  \BibitemOpen
  \bibfield  {author} {\bibinfo {author} {\bibfnamefont {R.}~\bibnamefont
  {Casadio}}, \bibinfo {author} {\bibfnamefont {A.}~\bibnamefont
  {Kamenshchik}}, \ and\ \bibinfo {author} {\bibfnamefont {J.}~\bibnamefont
  {Ovalle}},\ }\href {\doibase 10.1103/PhysRevD.110.044001} {\bibfield
  {journal} {\bibinfo  {journal} {Phys. Rev. D}\ }\textbf {\bibinfo {volume}
  {110}},\ \bibinfo {pages} {044001} (\bibinfo {year} {2024})}\BibitemShut
  {NoStop}%
\bibitem [{\citenamefont {Casadio}\ \emph {et~al.}(2025)\citenamefont
  {Casadio}, \citenamefont {Kamenshchik},\ and\ \citenamefont
  {Ovalle}}]{Casadio:2025pun}%
  \BibitemOpen
  \bibfield  {author} {\bibinfo {author} {\bibfnamefont {R.}~\bibnamefont
  {Casadio}}, \bibinfo {author} {\bibfnamefont {A.}~\bibnamefont
  {Kamenshchik}}, \ and\ \bibinfo {author} {\bibfnamefont {J.}~\bibnamefont
  {Ovalle}},\ }\href {\doibase 10.1103/PhysRevD.111.064036} {\bibfield
  {journal} {\bibinfo  {journal} {Phys. Rev. D}\ }\textbf {\bibinfo {volume}
  {111}},\ \bibinfo {pages} {064036} (\bibinfo {year} {2025})}\BibitemShut
  {NoStop}%
\bibitem [{\citenamefont {Lukash}\ and\ \citenamefont
  {Strokov}(2013)}]{Lukash:2013ts}%
  \BibitemOpen
  \bibfield  {author} {\bibinfo {author} {\bibfnamefont {V.~N.}\ \bibnamefont
  {Lukash}}\ and\ \bibinfo {author} {\bibfnamefont {V.~N.}\ \bibnamefont
  {Strokov}},\ }\href {\doibase 10.1142/S0217751X13500073} {\bibfield
  {journal} {\bibinfo  {journal} {Int. J. Mod. Phys. A}\ }\textbf {\bibinfo
  {volume} {28}},\ \bibinfo {pages} {1350007} (\bibinfo {year} {2013})},\
  \Eprint {http://arxiv.org/abs/1301.5544} {arXiv:1301.5544 [gr-qc]}
  \BibitemShut {NoStop}%
\bibitem [{\citenamefont {Ovalle}(2023)}]{Ovalle:2023vvu}%
  \BibitemOpen
  \bibfield  {author} {\bibinfo {author} {\bibfnamefont {J.}~\bibnamefont
  {Ovalle}},\ }\href {\doibase 10.1103/PhysRevD.107.104005} {\bibfield
  {journal} {\bibinfo  {journal} {Phys. Rev. D}\ }\textbf {\bibinfo {volume}
  {107}},\ \bibinfo {pages} {104005} (\bibinfo {year} {2023})}\BibitemShut
  {NoStop}%
\bibitem [{\citenamefont {Arrechea}\ \emph {et~al.}(2025)\citenamefont
  {Arrechea}, \citenamefont {Liberati}, \citenamefont {Neshat},\ and\
  \citenamefont {Vellucci}}]{Arrechea:2025fkk}%
  \BibitemOpen
  \bibfield  {author} {\bibinfo {author} {\bibfnamefont {J.}~\bibnamefont
  {Arrechea}}, \bibinfo {author} {\bibfnamefont {S.}~\bibnamefont {Liberati}},
  \bibinfo {author} {\bibfnamefont {H.}~\bibnamefont {Neshat}}, \ and\ \bibinfo
  {author} {\bibfnamefont {V.}~\bibnamefont {Vellucci}},\ }\href {\doibase
  10.1103/wk7j-yg1t} {\bibfield  {journal} {\bibinfo  {journal} {Phys. Rev. D}\
  }\textbf {\bibinfo {volume} {112}},\ \bibinfo {pages} {044024} (\bibinfo
  {year} {2025})}\BibitemShut {NoStop}%
\bibitem [{\citenamefont {Schwarzschild}(1916)}]{schwarzschild1916b}%
  \BibitemOpen
  \bibfield  {author} {\bibinfo {author} {\bibfnamefont {K.}~\bibnamefont
  {Schwarzschild}},\ }\href@noop {} {\bibfield  {journal} {\bibinfo  {journal}
  {Sitzungsber. Preuss. Akad. Wiss. Berlin (Math. Phys.)}\ }\textbf {\bibinfo
  {volume} {1916}},\ \bibinfo {pages} {424} (\bibinfo {year}
  {1916})}\BibitemShut {NoStop}%
\bibitem [{\citenamefont {Aoki}\ \emph
  {et~al.}(2021{\natexlab{a}})\citenamefont {Aoki}, \citenamefont {Onogi},\
  and\ \citenamefont {Yokoyama}}]{Aoki:2020nzm}%
  \BibitemOpen
  \bibfield  {author} {\bibinfo {author} {\bibfnamefont {S.}~\bibnamefont
  {Aoki}}, \bibinfo {author} {\bibfnamefont {T.}~\bibnamefont {Onogi}}, \ and\
  \bibinfo {author} {\bibfnamefont {S.}~\bibnamefont {Yokoyama}},\ }\href
  {\doibase 10.1142/S0217751X21502018} {\bibfield  {journal} {\bibinfo
  {journal} {Int. J. Mod. Phys. A}\ }\textbf {\bibinfo {volume} {36}},\
  \bibinfo {pages} {2150201} (\bibinfo {year} {2021}{\natexlab{a}})},\ \Eprint
  {http://arxiv.org/abs/2010.07660} {arXiv:2010.07660 [gr-qc]} \BibitemShut
  {NoStop}%
\bibitem [{\citenamefont {Aoki}\ and\ \citenamefont
  {Onogi}(2022)}]{Aoki:2022gez}%
  \BibitemOpen
  \bibfield  {author} {\bibinfo {author} {\bibfnamefont {S.}~\bibnamefont
  {Aoki}}\ and\ \bibinfo {author} {\bibfnamefont {T.}~\bibnamefont {Onogi}},\
  }\href {\doibase 10.1142/S0217751X22501299} {\bibfield  {journal} {\bibinfo
  {journal} {Int. J. Mod. Phys. A}\ }\textbf {\bibinfo {volume} {37}},\
  \bibinfo {pages} {2250129} (\bibinfo {year} {2022})},\ \Eprint
  {http://arxiv.org/abs/2201.09557} {arXiv:2201.09557 [hep-th]} \BibitemShut
  {NoStop}%
\bibitem [{\citenamefont {Aoki}\ \emph
  {et~al.}(2021{\natexlab{b}})\citenamefont {Aoki}, \citenamefont {Onogi},\
  and\ \citenamefont {Yokoyama}}]{Aoki:2020prb}%
  \BibitemOpen
  \bibfield  {author} {\bibinfo {author} {\bibfnamefont {S.}~\bibnamefont
  {Aoki}}, \bibinfo {author} {\bibfnamefont {T.}~\bibnamefont {Onogi}}, \ and\
  \bibinfo {author} {\bibfnamefont {S.}~\bibnamefont {Yokoyama}},\ }\href
  {\doibase 10.1142/S0217751X21500986} {\bibfield  {journal} {\bibinfo
  {journal} {Int. J. Mod. Phys. A}\ }\textbf {\bibinfo {volume} {36}},\
  \bibinfo {pages} {2150098} (\bibinfo {year} {2021}{\natexlab{b}})},\ \Eprint
  {http://arxiv.org/abs/2005.13233} {arXiv:2005.13233 [gr-qc]} \BibitemShut
  {NoStop}%
\bibitem [{\citenamefont {Wheeler}(2022)}]{Wheeler:2022xdo}%
  \BibitemOpen
  \bibfield  {author} {\bibinfo {author} {\bibfnamefont {J.}~\bibnamefont
  {Wheeler}},\ }\href {\doibase 10.1088/1361-6382/ac89c7} {\bibfield  {journal}
  {\bibinfo  {journal} {Class. Quant. Grav.}\ }\textbf {\bibinfo {volume}
  {39}},\ \bibinfo {pages} {197001} (\bibinfo {year} {2022})},\ \Eprint
  {http://arxiv.org/abs/2205.11639} {arXiv:2205.11639 [gr-qc]} \BibitemShut
  {NoStop}%
\bibitem [{\citenamefont {Bengtsson}\ \emph {et~al.}(2013)\citenamefont
  {Bengtsson}, \citenamefont {Jakobsson},\ and\ \citenamefont
  {Senovilla}}]{Bengtsson:2013hla}%
  \BibitemOpen
  \bibfield  {author} {\bibinfo {author} {\bibfnamefont {I.}~\bibnamefont
  {Bengtsson}}, \bibinfo {author} {\bibfnamefont {E.}~\bibnamefont
  {Jakobsson}}, \ and\ \bibinfo {author} {\bibfnamefont {J.~M.~M.}\
  \bibnamefont {Senovilla}},\ }\href {\doibase 10.1103/PhysRevD.88.064012}
  {\bibfield  {journal} {\bibinfo  {journal} {Phys. Rev. D}\ }\textbf {\bibinfo
  {volume} {88}},\ \bibinfo {pages} {064012} (\bibinfo {year} {2013})},\
  \Eprint {http://arxiv.org/abs/1306.6486} {arXiv:1306.6486 [gr-qc]}
  \BibitemShut {NoStop}%
\bibitem [{\citenamefont {Plebanski}(1964)}]{Plebanski:1964}%
  \BibitemOpen
  \bibfield  {author} {\bibinfo {author} {\bibfnamefont {J.}~\bibnamefont
  {Plebanski}},\ }\href@noop {} {\bibfield  {journal} {\bibinfo  {journal}
  {Acta Phys. Pol.}\ }\textbf {\bibinfo {volume} {26}},\ \bibinfo {pages} {963}
  (\bibinfo {year} {1964})}\BibitemShut {NoStop}%
\bibitem [{\citenamefont {Hawking}\ and\ \citenamefont
  {Ellis}(2011)}]{Hawking:1973uf}%
  \BibitemOpen
  \bibfield  {author} {\bibinfo {author} {\bibfnamefont {S.~W.}\ \bibnamefont
  {Hawking}}\ and\ \bibinfo {author} {\bibfnamefont {G.~F.~R.}\ \bibnamefont
  {Ellis}},\ }\href {\doibase 10.1017/CBO9780511524646} {\emph {\bibinfo
  {title} {{The Large Scale Structure of Space-Time}}}},\ Cambridge Monographs
  on Mathematical Physics\ (\bibinfo  {publisher} {Cambridge University
  Press},\ \bibinfo {year} {2011})\BibitemShut {NoStop}%
\bibitem [{\citenamefont {Martin-Moruno}\ and\ \citenamefont
  {Visser}(2017)}]{Martin-Moruno:2017exc}%
  \BibitemOpen
  \bibfield  {author} {\bibinfo {author} {\bibfnamefont {P.}~\bibnamefont
  {Martin-Moruno}}\ and\ \bibinfo {author} {\bibfnamefont {M.}~\bibnamefont
  {Visser}},\ }\href {\doibase 10.1007/978-3-319-55182-1_9} {\bibfield
  {journal} {\bibinfo  {journal} {Fundam. Theor. Phys.}\ }\textbf {\bibinfo
  {volume} {189}},\ \bibinfo {pages} {193} (\bibinfo {year} {2017})},\ \Eprint
  {http://arxiv.org/abs/1702.05915} {arXiv:1702.05915 [gr-qc]} \BibitemShut
  {NoStop}%
\bibitem [{\citenamefont {Maeda}\ and\ \citenamefont
  {Martinez}(2020)}]{Maeda:2018hqu}%
  \BibitemOpen
  \bibfield  {author} {\bibinfo {author} {\bibfnamefont {H.}~\bibnamefont
  {Maeda}}\ and\ \bibinfo {author} {\bibfnamefont {C.}~\bibnamefont
  {Martinez}},\ }\href {\doibase 10.1093/ptep/ptaa009} {\bibfield  {journal}
  {\bibinfo  {journal} {PTEP}\ }\textbf {\bibinfo {volume} {2020}},\ \bibinfo
  {pages} {043E02} (\bibinfo {year} {2020})},\ \Eprint
  {http://arxiv.org/abs/1810.02487} {arXiv:1810.02487 [gr-qc]} \BibitemShut
  {NoStop}%
\bibitem [{\citenamefont {Maeda}\ and\ \citenamefont
  {Harada}(2022)}]{Maeda:2022vld}%
  \BibitemOpen
  \bibfield  {author} {\bibinfo {author} {\bibfnamefont {H.}~\bibnamefont
  {Maeda}}\ and\ \bibinfo {author} {\bibfnamefont {T.}~\bibnamefont {Harada}},\
  }\href {\doibase 10.1088/1361-6382/ac8861} {\bibfield  {journal} {\bibinfo
  {journal} {Class. Quant. Grav.}\ }\textbf {\bibinfo {volume} {39}},\ \bibinfo
  {pages} {195002} (\bibinfo {year} {2022})},\ \Eprint
  {http://arxiv.org/abs/2205.12993} {arXiv:2205.12993 [gr-qc]} \BibitemShut
  {NoStop}%
\bibitem [{\citenamefont {Wang}\ and\ \citenamefont
  {Battista}(2026{\natexlab{a}})}]{Wang:2026jvo}%
  \BibitemOpen
  \bibfield  {author} {\bibinfo {author} {\bibfnamefont {Z.-L.}\ \bibnamefont
  {Wang}}\ and\ \bibinfo {author} {\bibfnamefont {E.}~\bibnamefont
  {Battista}},\ }\href@noop {} {\  (\bibinfo {year} {2026}{\natexlab{a}})},\
  \Eprint {http://arxiv.org/abs/2604.16545} {arXiv:2604.16545 [gr-qc]}
  \BibitemShut {NoStop}%
\bibitem [{\citenamefont {Wang}\ and\ \citenamefont
  {Battista}(2026{\natexlab{b}})}]{Wang:2026sqr}%
  \BibitemOpen
  \bibfield  {author} {\bibinfo {author} {\bibfnamefont {Z.-L.}\ \bibnamefont
  {Wang}}\ and\ \bibinfo {author} {\bibfnamefont {E.}~\bibnamefont
  {Battista}},\ }\href@noop {} {\  (\bibinfo {year} {2026}{\natexlab{b}})},\
  \Eprint {http://arxiv.org/abs/2605.03428} {arXiv:2605.03428 [gr-qc]}
  \BibitemShut {NoStop}%
\end{thebibliography}%
\bibliographystyle{apsrev4-1}
%
%
\end{document}